\documentclass[twocolumn, prl, longbibliography]{revtex4-2}
\usepackage{graphicx, amsmath}

\def\rvec{\mathbf{r}}

\def\bphi{\bar{\phi}_{\rm gb}}

\def\phiGbGb{\langle \phi_{\rm gb} \rangle_{\rm gb}}
\def\phiLc{\langle \phi_{\rm lc} \rangle_{\rm lc}}
\def\phiGbLc{\langle \phi_{\rm gb} \rangle_{\rm lc}}
\def\Lol{L/\ell} 
\def\Var{\mathrm{Var}}

\usepackage[colorlinks=true,
	pdfstartview = FitV,
	linkcolor    = linkcolor,
	citecolor    = linkcolor,
	urlcolor     = linkcolor,	
	hyperindex   = true,
	hyperfigures = false]{hyperref}
\usepackage{color}
\definecolor{linkcolor}{rgb}{0,0,0.6}

\begin{document} 

\title{Biased ensembles of pulsating active matter}

\begin{abstract}
We discover unexpected connections between packing configurations and rare fluctuations in dense systems of active particles subject to pulsation of size. Using large deviation theory, we examine biased ensembles which select atypical realizations of the dynamics exhibiting high synchronization in particle size. We show that the order emerging at high bias can manifest as distinct dynamical states featuring high to vanishing pulsation current. Remarkably, transitions between these states arise from changing the system geometry at fixed bias and constant density. We rationalize such transitions as arising from the change in packing configurations which, depending on box geometry, may induce highly ordered or geometrically frustrated structures. Furthermore, we reveal that a master curve in the unbiased dynamics, correlating polydispersity and current, helps predict the dynamical state emerging in the biased dynamics. Finally, we demonstrate that deformation waves can propagate under suitable geometries when biasing with local order.
\end{abstract}

\author{William D. Pi\~neros}
\author{\'Etienne Fodor}
\affiliation{Department of Physics and Materials Science, University of Luxembourg, L-1511 Luxembourg, Luxembourg}
    
\maketitle


\textit{Introduction.}---Active matter encompasses systems which constantly dissipate energy to sustain collective behaviors far from equilibrium. For instance, assemblies of self-propelled particles (SPPs)~\cite{Marchetti2013, Bechinger2016, Marchetti2018} yield nonequilibrium phenomena which have been extensively studied, such as a polarized collective motion~\cite{Vicsek1995, Chate2020} and a phase separation without attractive interactions~\cite{MIPS1, MIPSrev}. Energy dissipation can also take other forms beyond motility, opening the door to novel physics beyond that of SPPs. For instance, in some biological tissues (e.g., epithelial~\cite{Lecuit2022}, cardiac~\cite{Karma2013}, and uterine~\cite{Elad2017} tissues), each cell can sustain periodic changes of shape, leading to the propagation of deformation waves. A recent model of pulsating active matter (PAM) has captured such waves in terms of densely packed particles whose sizes constantly pulsate~\cite{PulsatingAM}. In contrast with other models of deforming particles where waves have not been observed~\cite{Tjhung2016, Tjhung2017, Oyama2019, Koyano2019}, PAM relies on synchronizing nearby sizes~\cite{Togashi2019, PulsatingAM, Manacorda2024, Liu2024, Li2024}. In the absence of any synchronizing interaction, it is largely unclear under which conditions deformation waves could still potentially emerge.

Biased ensembles (BEs) offer the perfect toolbox~\cite{Jack2020} to search for waves in non-synchronizing PAM. Building on large deviation theory~\cite{Touchette2009}, BEs select the rare trajectories which (i)~achieve some atypical statistics of a chosen observable, while (ii)~deviating the least from the original, unbiased dynamics. In practice, BEs do not presume {\it how} the system should accommodate the constraints (i-ii). At a sufficiently large bias, the dynamics are given enough play to explore novel configurations, potentially yielding dynamical phase transitions~\cite{Wijland2007, Hurtado2017, glassStructureLD}. Rare trajectories can actually be mapped into an effective dynamics~\cite{Chetrite2015, Jack2015, Limmer2018, Suri2019}, which constitute the optimal mechanism for stabilizing the phases selected by BEs. This connection between optimal control and large deviations has inspired novel strategies for material design~\cite{Limmer2023, CytoskeletonLD}.

In active matter, BEs have already been used to unravel novel mechanisms for promoting collective effects~\cite{Jack2022}. For instance, in large deviations of SPPs~\cite{ABPLD0, Whitelam2018, GrandPre2021}, BEs have revealed that alignment effectively emerges from avoiding collisions between nearby particles, yielding collective motion~\cite{ABPLD1, ABPLD2, Agranov2023}. 
These results are reminiscent of the collective behaviors emerging in some models of SPPs without alignment~\cite{alignment_nosync1, alignment_ring_nosync2, flocking_noalignment}, and stand in contrast with standard flocking models which require ad hoc aligning rules~\cite{Chate2020}. In a similar fashion, it is tempting to examine whether BEs of non-synchronizing PAM entail unexpected transitions, potentially uncovering novel pathways towards wave formation.

\begin{figure}[b]
	\centering
	\includegraphics[scale=.3]{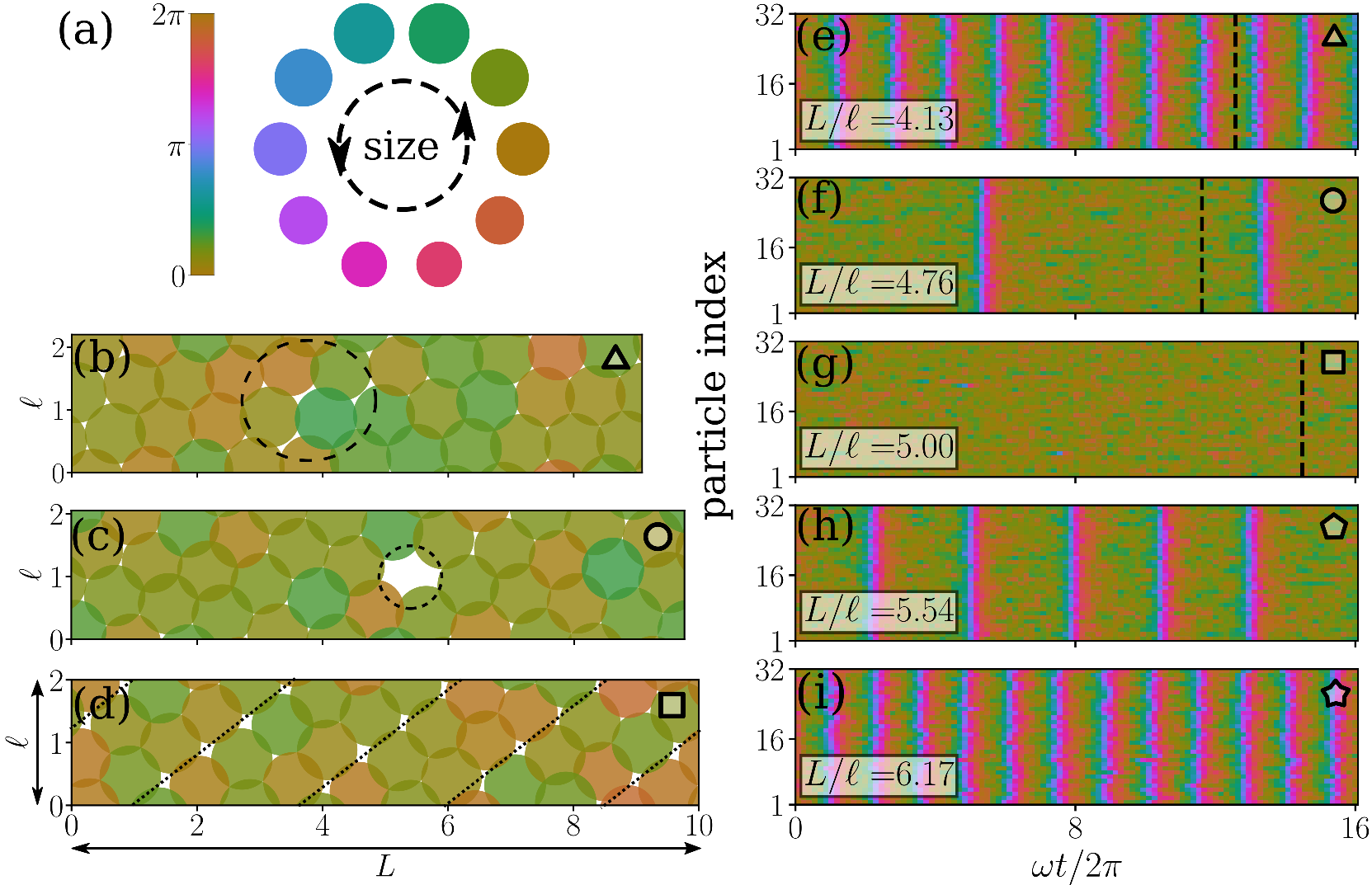}
 	\caption{
  (a)~Size of particles subject to a periodic pulsation controlled by an internal phase. 
  In biased ensembles at high density, changing the box aspect ratio $\Lol$ with sides $\{\ell, L\}$ results in various configurations with either (b)~defects, (c)~voids, or (d)~regular structures. These correspond, respectively, to varying collective states for different $\Lol$: (e,i)~cycles with periodic size change, (f,h)~intermittent behavior with aperiodic size change, and (g)~arrest where particle repulsion yields a uniform frozen size. Particles are indexed per increasing position along the $L$-axis. Dashed lines refer to the snapshots shown in (b)-(d) and symbols indicate parameter values in Fig.~\ref{fig:phase_diagram}.
	}
\label{fig:N32_trajs}
\end{figure}

In this letter, we investigate the collective dynamics in BEs of dense assemblies of pulsating particles in a two-dimensional box [Fig.~\ref{fig:N32_trajs}]. Starting from configurations without any synchronization, we reveal how biasing with a global order parameter promotes transitions towards various homogeneous states: cycles, arrest, and intermittent behavior. Remarkably, the packing constraint imposed by the box geometry, which restricts the unbiased configurations and hence its fluctuations, also selects for one of the three ordered states under bias. Specifically, we show that slight variations in the unbiased statistics of some relevant observables allows one to anticipate the emergence of transitions in the biased dynamics. Finally, we reveal that deformation waves can be stabilized for specific geometries when biasing with a local order parameter. Overall, our results demonstrate that, at fixed bias and constant density, controlling the box geometry is a novel route towards unexpected phase transitions in BEs of PAM.


\textit{Pulsating active matter: The role of box geometry.}---We consider a two-dimensional system of pulsating particles [Fig.~\ref{fig:N32_trajs}] whose sizes change as
\begin{equation}\label{eq:sigma}
	\sigma_i = \frac{\sigma_0}{2} \frac{1+\lambda \sin \theta_i}{1+\lambda},
\end{equation}
where $\sigma_0=1$ is the base size, $\theta_i$ the internal phase of particle $i$, and $\lambda=0.05$ the pulsation amplitude. The particles follow overdamped Langevin dynamics:
\begin{align}
	\dot{\rvec}_i &= -\mu_r \partial_{\rvec_i} V +\sqrt{2D_r} \boldsymbol\xi_i ,
	\label{eq:dynamics_0}
	\\
	\dot\theta_i &= \omega - \mu_\theta \partial_{\theta_i} V + \sqrt{2D_\theta}\eta_i .
	\label{eq:dynamics} 
\end{align}
The potential $V=\sum_{i,j<i} U(a_{ij})$ depends on the scaled distance $a_{ij} = |\rvec_j-\rvec_i|/(\sigma_i+\sigma_j)$, and $(\boldsymbol\xi_i, \eta_i)$ are uncorrelated Gaussian white noises with unit correlations. The phase drift is constant and set to $\omega=10$.  The diffusion coefficients $D_{r/\theta}$ and mobilities $\mu_{r/\theta}$ follow the fluctuation-dissipation relation ($D_{r}/\mu_{r}=D_{\theta}/\mu_{\theta}=T$) and are set to unity. Then, in the absence of drift ($\omega=0$), the system follows an equilibrium dynamics which relaxes to the Boltzmann distribution $P\sim e^{-V/T}$. Interactions follow volume exclusion via a Weeks-Chandler-Anderson potential $U(a) = U_0 (a^{-12}- 2 a^{-6})$ with $U_0=1$ and cut-off set at $a=1$. In contrast with~\cite{PulsatingAM}, we do not consider here any synchronizing interaction between phases. In what follows, we run simulations with periodic boundary conditions for $N=32$ particles at density $\rho=1.6$ (unless stated otherwise), and we examine how the box aspect ratio $\Lol$ [Fig.~\ref{fig:N32_trajs}] impacts the emerging dynamics.

\begin{figure}
	\centering
	\includegraphics[scale=0.23]{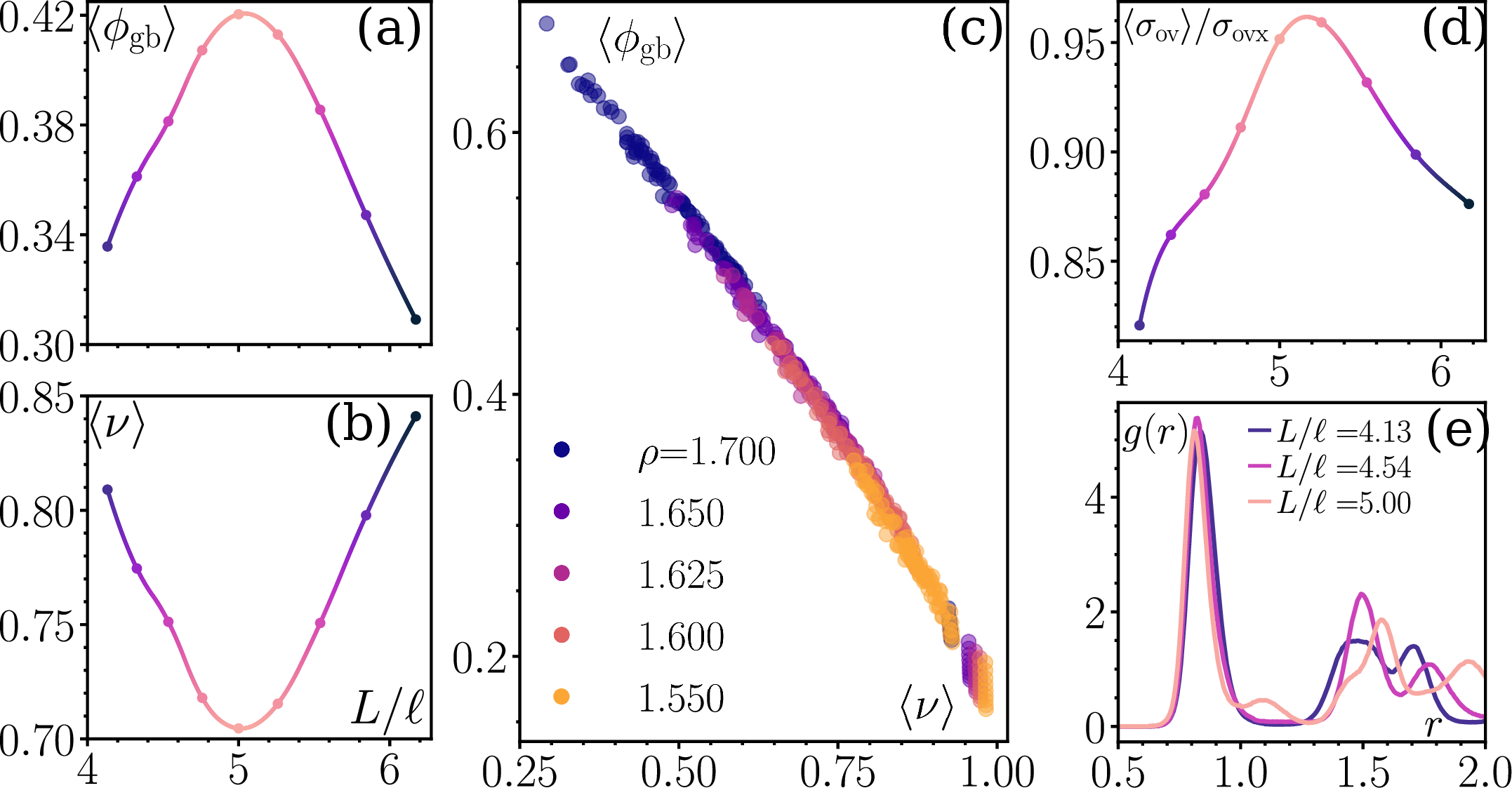}
	\caption{(a-b)~Global order $\langle\phi_{\rm gb}\rangle$ and phase current $\langle\nu\rangle$ as functions of the aspect ratio $L/\ell$. Lines are guides to the eye.
 	(c)~Correlation between global order $\langle\phi_{\rm gb}\rangle$ and phase current $\langle\nu\rangle$ for various densities $\rho=N/(L\ell)$, changing both $L/\ell$ and the particle number $N$.
  	(d)~Average overlap $\langle\sigma_{\rm ov}\rangle$, relative to that of hexagonal packing $\sigma_{\rm ovx}$ at same density, for varying values of $\Lol$. (e)~Radial distribution $g$ as a function of the interparticle distance $r$. Line colors in (b),(d-e) as per the value of $\langle \phi_{\rm gb} \rangle$ in (a). 
	}
\label{fig:templating}
\end{figure}

We start by evaluating the global order parameter quantifying phase synchronization across the system:
\begin{equation}\label{eq:pgb}
	\phi_{\rm gb} = \frac{1}{N}\Bigg| \sum_{j=1}^N e^{i\theta_j} \Bigg | = \frac{1}{N} \sqrt{ \sum_{i,j=1}^N \cos(\theta_i-\theta_j) } .
\end{equation}
Configurations with a nearly uniform size distribution have $\phi_{\rm gb}\simeq1$, whereas those with high polydispersity have $\phi_{\rm gb}\simeq0$. Despite the absence of any synchronizing interaction, our simulations exhibit a moderate ordering of particle sizes [Fig.~\ref{fig:templating}(a)]. Indeed, the repulsion term in Eq.~\eqref{eq:dynamics} constrains the sizes to fluctuate around a preferred value. This effect is captured at mean-field level by approximating $\partial_{\theta_i}V\approx(\partial_\varphi V) (\partial_{\theta_i}\varphi)$~\cite{PulsatingAM}. The coefficient $\partial_\varphi V$ increases with $\rho$, and the packing fraction $\varphi = (\pi\rho/N) \sum_i \sigma_i^2$ admits the same local minimum for each $\theta_i\in[0,2\pi]$, thus favoring order at high $\rho$.

Interestingly, we observe that $\langle\phi_{\rm gb}\rangle$ (where $\langle\cdot\rangle$ indicates average over realizations) strongly varies with the aspect ratio $\Lol$, with a maximum at $\Lol\simeq 5$ [Fig.~\ref{fig:templating}(a)], showing that one can actually enhance order by appropriately tuning the box geometry. We find that such a variation with $\Lol$ wanes gradually for larger $N$ at fixed $\rho$, but remains noticeable even for systems with $N\simeq 10^2$ (see Fig.~S2 in~\cite{SM}). The enhancement of order is accompanied by a dynamical slowdown, as indicated by the reduction of the average phase current [Fig.~\ref{fig:templating}(b)]:
\begin{align}
  \nu = \frac{1}{N \omega t}\sum_{i=1}^N \int^t_0 dt' \dot{\theta}_i(t') .
\end{align}
Similar results hold for other box ratios and particle numbers, see Fig.~S5 in~\cite{SM}. In fact, plotting $\langle\phi_{\rm gb}\rangle$ against $\langle\nu\rangle$ for different values of $N$, $\Lol$, and $\rho$ results in a master curve, where increasing order systematically correlates with decreasing current [Fig.~\ref{fig:templating}(c)]. Interestingly, some of the curves, shown in Fig.~\ref{fig:templating}(c) for different $\rho$, overlap. Indeed, for some combinations of $N$ and $\Lol$, systems at various $\rho$ can experience an equivalent average repulsion, yielding a similar set of values for $\langle\phi_{\rm gb}\rangle$ and $\langle\nu\rangle$.

We can rationalize these results from a packing perspective. To this end, we explore how repulsion varies for different values of $\Lol$ by computing the average overlap:
\begin{equation}
    \langle \sigma_\mathrm{ov} \rangle = \frac{2}{N(N-1)}\sum_{i,j\in \partial_i} \langle \sigma_i+\sigma_j - |{\bf r}_i-{\bf r}_j|\rangle ,
\end{equation}
where the sum runs over particles for which $\sigma_i+\sigma_j > |{\bf r}_i-{\bf r}_j|$. We find the average overlap peaks at $\Lol\simeq 5$ [Fig.~\ref{fig:templating}(d)], which coincides with the maximum of $\langle\phi_{\rm gb}\rangle$ and the minimum of $\langle\nu\rangle$ [Figs.~\ref{fig:templating}(a-b)]. 
Interestingly, $\langle\sigma_\mathrm{ov}\rangle$ is comparable to the overlap $\sigma_\mathrm{ovx}$ of a hexagonal packing at the same density [Appendix~A], and a regular structure can be identified at $\Lol\simeq5$; see Sec.~I.A in~\cite{SM}.  
These results indicate that higher overlap (hence repulsion) induce more regular arrangements, corresponding to lower polydispersity and lower pulsation current in the system. To illustrate further the relation between structure and box geometry, we evaluate the radial distribution function (rdf)
\begin{equation}
    g(r) = \frac 1 \rho \sum_{i,j\neq i}\langle\delta(r-|{\bf r}_i-{\bf r}_j|)\rangle
\end{equation}
and observe that indeed density correlations change drastically for varying $\Lol$ [Fig.~\ref{fig:templating}(e)]. In particular, the rdf for $\Lol=5$ features a secondary peak at $r\simeq 1.1$ not present for other values of $\Lol$, which hints at qualitative differences between the corresponding packing structures. A similar dependence of structure on box geometry has been reported in systems of monodisperse, passive particles~\cite{CrystalAssemblyBoxEffects, TorquatoCrystalAssembly, OpenSelfAssembly}.

In short, our findings show that changing the box geometry alters the packing structure assumed by pulsating particles, which in turn provides a route to controlling order and current at fixed density.


\textit{Ensembles biased by global order: Cycles and arrest.}---We study the large deviations of the dynamics with respect to $\phi_{\rm gb}$ [Eq.~\eqref{eq:pgb}]. In particular, we seek trajectories for which the time average
\begin{equation}
	\bar\phi_{\rm gb} = \frac{1}{t_o}\int_0^{t_o} \phi_{\rm gb}(t) dt
\end{equation}
displays atypically large values at large observation time $t_o$. To this end, we use a BE selecting for such trajectories through rare realizations of the noise terms in Eqs.~\eqref{eq:dynamics_0} and~\eqref{eq:dynamics}. Chiefly, we denote averages with respect to this BE as
\begin{equation}\label{eq:sgb}
	\langle \cdot \rangle_{\rm gb} = \frac{\langle \cdot \, e^{- s N t_o\bar\phi_{\rm gb}} \rangle}{\langle e^{-s N t_o\bar\phi_{\rm gb}} \rangle} .
\end{equation}
The bias strength $s$ effectively controls the statistics of $\bphi$. At vanishing bias, $s=0$, one recovers the ensemble of the original dynamics: $\langle\cdot\rangle_{\rm gb} = \langle\cdot\rangle$. In this work, we implement the trajectory selection via a cloning algorithm using population dynamics~\cite{CloningMethod}. It consists in simulating $n_c$ identical (though distinctly seeded) parallel runs, which are regularly replicated/pruned through a sampling procedure parametrized by $s$. In the limit of large $n_c$ and large $t_o$, this procedure converges to a BE whose trajectories represent the least unlikely dynamics to stabilize the desired atypical statistics of $\bphi$.

For large $N$, numerical convergence becomes increasingly challenging~\cite{Jack2020}. Generally, a useful method consists in adding some terms in the dynamics which effectively approximate trajectory selection and improves convergence~\cite{Nemoto2016, ABPLD1, Jack2020, ABPLD2, TimeCrystalsKuramotoLD}. Here, we consider fully-connected synchronizing interactions:
\begin{equation}
	\dot\theta_i = \omega - \mu_\theta \partial_{\theta_i} V + \varepsilon \sum_{j=1}^N \sin(\theta_j-\theta_i) + \sqrt{2D_\theta}\eta_i .
	\label{eq:mod_dynamics}
\end{equation}
We adapt our numerical selection of trajectories~\cite{SM} to ensure that, while using the dynamics in Eq.~\eqref{eq:mod_dynamics}, we still sample the proper BE [Eq.~\eqref{eq:sgb}] defined independently of synchronizing interactions. Furthermore, for each run, we heuristically adjust the amplitude $\varepsilon$ throughout the trajectory~\cite{SM}, which converges at large $t_0$ to a value determined by $s$. In what follows, we are interested in the regime of bias which promotes order, namely atypically large $\bphi$, corresponding here to $s<0$ and $\varepsilon>0$.

\begin{figure}
	\centering
	\includegraphics[scale=.31]{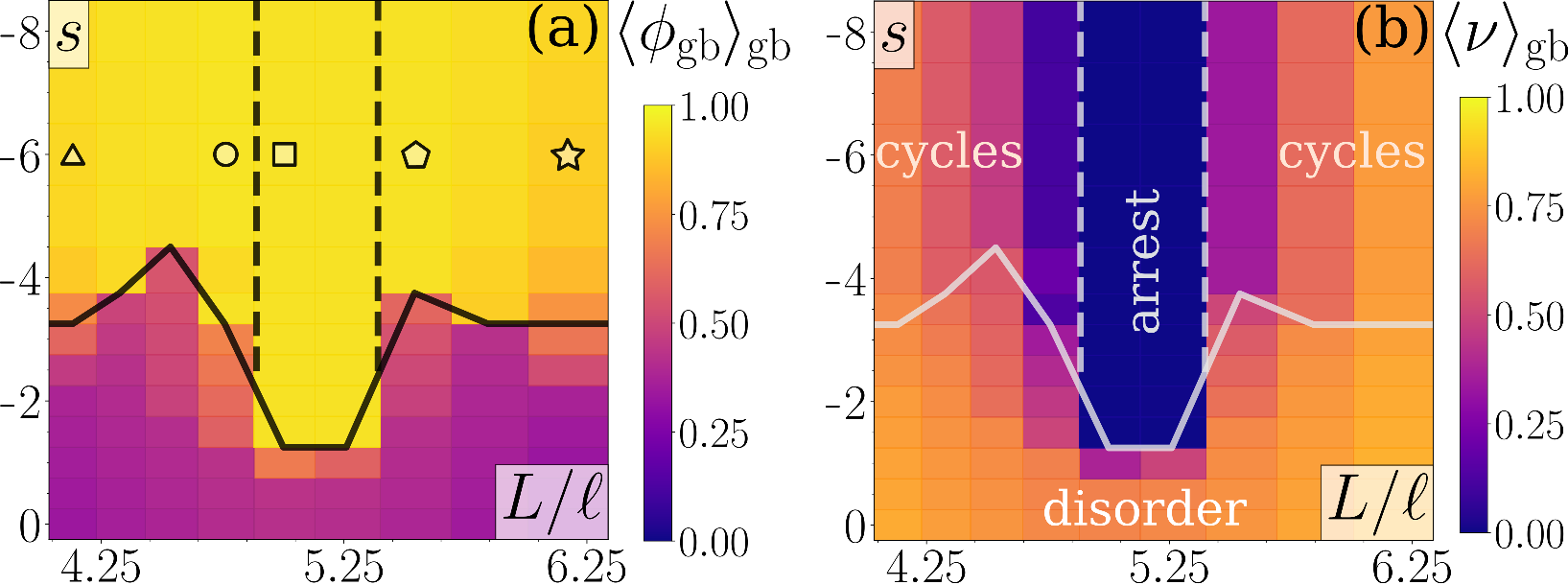}
	\caption{Phase diagram in ensembles biased by global order [Eq.~\eqref{eq:sgb}] in terms of the bias strength $s$ and the aspect ratio $L/\ell$: (a)~global order $\phiGbGb$, and (b)~phase current $\langle\nu\rangle_{\rm gb}$. Boundary lines are for $\phiGbGb = 0.65$ (solid) and $\langle\nu\rangle_{\rm gb} = 0.1$ (dashed). Markers refer to various trajectories at $s=-6$, as shown in Fig.~\ref{fig:N32_trajs}.}
\label{fig:phase_diagram}
\end{figure}

Starting from the unbiased configuration with highest $\langle\phi_{\rm gb}\rangle$ and lowest $\langle\nu\rangle$, namely for $\Lol\simeq 5$ [Figs.~\ref{fig:templating}(a) and~\ref{fig:templating}(b)], increasing $|s|$ yields a highly ordered state without phase current: $\langle\phi_{\rm gb}\rangle_{\rm gb} \simeq 1$ and $\langle\nu\rangle_{\rm gb}\simeq 0$ [Figs.~\ref{fig:phase_diagram}(a) and~\ref{fig:phase_diagram}(b)]. Such a configuration is analogous to the arrested state previously reported in synchronizing PAM~\cite{PulsatingAM, Manacorda2023}. Remarkably, considering an unbiased configuration at the tail of the curve $\langle\nu\rangle$ vs $\Lol$ [Fig.~\ref{fig:templating}(b)], increasing $|s|$ now yields an ordered state with non-vanishing averaged current [Figs.~\ref{fig:phase_diagram}(a) and~\ref{fig:phase_diagram}(b)], which is reminiscent of the cycling state in synchronizing PAM~\cite{PulsatingAM, Manacorda2023}. Therefore, the box geometry not only impacts the unbiased dynamics of non-synchronizing PAM, it also strongly influences its rare fluctuations, yielding two different types of ordered states.

Varying $\Lol$ at constant $|s|>4.5$, the transition between cycling and arrest [Fig.~\ref{fig:phase_diagram}(b)] mirrors the slowdown of the unbiased dynamics [Fig.~\ref{fig:templating}(b)]. Again, this result can be rationalized from a packing perspective. Specifically, arrest is associated with a regular packing [Fig.~\ref{fig:N32_trajs}(d)] which impedes the periodic expansion and contraction of particles [Fig.~\ref{fig:N32_trajs}(g)], whereas cycles have a defective packing [Fig.~\ref{fig:N32_trajs}(b)] which facilitates global changes in particle sizes [Figs.~\ref{fig:N32_trajs}(e) and~\ref{fig:N32_trajs}(i)]. Note that $\langle\phi_{\rm gb} \rangle_{\rm gb}$ is slightly higher for arrest compared with cycles [Fig.~\ref{fig:phase_diagram}(a)], so that regular packing is associated with a reduced polydispersity, as in the unbiased case [Fig.~\ref{fig:templating}]. Interestingly, we also observe an intermittent dynamics with aperiodic size changes [Figs.~\ref{fig:N32_trajs}(f) and~\ref{fig:N32_trajs}(h)] whose packing contains motile voids [Fig.~\ref{fig:N32_trajs}(c)]. Indeed, measurements of structural order clearly highlight structural differences between each one of these dynamical states; see Appendix~B, and Fig.~S4 in~\cite{SM}.

Overall, our results for $(N,\rho)=(32,1.6)$ show that packing configurations, imposed by the box geometry, impact both unbiased and biased dynamics. Importantly, we reveal that the unbiased statistics actually allows one to anticipate how the system orders as a function of $\Lol$ in our BE. We find a similar effect is generically observed for other values of $(N,\rho)$; for instance, see the phase diagram for $(N,\rho)=(26,1.6)$ in Fig.~S5 of~\cite{SM}.


\textit{Ensembles biased by local order: Deformation waves.}---In synchronizing PAM~\cite{PulsatingAM, Manacorda2023}, deformation waves emerge as a competition between arrest and cycling. Given that in non-synchronizing PAM the BE promoting global order [Eq.~\eqref{eq:sgb}] yields arrest and cycles [Fig.~\ref{fig:phase_diagram}], it is intriguing to understand what class of BE may also induce deformation waves. To this end, we introduce the local order parameter
\begin{equation}
	\phi_{\rm lc} = \frac{1}{N}\sum_{i=1}^N\sum_{j=1}^{n_i} \frac{\cos{(\theta_j-\theta_i)}}{n_i} ,
\end{equation}
where $n_i$ is the number of neighbors in contact with particle $i$, and the corresponding biased average
\begin{equation}\label{eq:slc}
	\langle \cdot \rangle_{\rm lc} = \frac{\langle \cdot \, e^{-s N t_o\bar \phi_{\rm lc}} \rangle}{\langle e^{-s N t_o \bar \phi_{\rm lc}} \rangle},
	\quad
	\bar\phi_{\rm lc} = \frac{1}{t_o}\int_0^{t_o} \phi_{\rm lc}(t) dt .
\end{equation}
To improve sampling, we now consider {\em locally} synchronizing interactions:
\begin{equation}
	\dot\theta_i = \omega - \mu_\theta \partial_{\theta_i} V + \varepsilon \sum_{j=1}^{n_i} \sin(\theta_j-\theta_i) + \sqrt{2D_\theta}\eta_i .
	\label{eq:mod_dynamics_2}
\end{equation}
In practice, Eq.~\eqref{eq:mod_dynamics_2} enhances convergence for the BE in Eq.~\eqref{eq:slc} at moderate $|s|$, while Eq.~\eqref{eq:mod_dynamics} actually works better for the same BE at large $|s|$. At each $s$, we systematically compare results obtained by employing either type of interaction (i.e., with local or global synchronization), and select the ones with optimal convergence~\cite{SM}.

\begin{figure}
	\centering
	\includegraphics[scale=1.13]{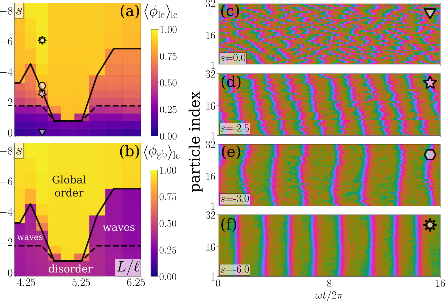}
	\caption{Phase diagram in ensembles biased by local order [Eq.~\eqref{eq:slc}] in terms of the bias strength $s$ and the aspect ratio $L/\ell$: (a)~local order $\phiLc$, and (b)~global order $\phiGbLc$. Boundary lines are for $\phiLc = 0.65$ (solid) and $\phiGbLc = 0.45$ (dashed). Markers refer to various trajectories at $\Lol=4.53$: (c)~disorder, (d,e)~waves, and (f)~cycles.
	}
\label{fig:phase_diagram_local}
\end{figure}

Interestingly, for values of $\Lol$ coincident with the minimum of $\langle\nu\rangle$ [Fig.~\ref{fig:templating}(b)], we observe again the emergence of an arrested state with local and global order [Figs.~\ref{fig:phase_diagram_local}(a) and~\ref{fig:phase_diagram_local}(b)] comparable to the results from the previous BE [Eq.~\eqref{fig:phase_diagram}]. 
In contrast, for $\Lol$ sufficiently far away from the minimum of $\langle\nu\rangle$, phase ordering now occurs through two distinct states. As $|s|$ increases, local order increases with negligible change in global order, i.e. $\langle\phi_{\rm lc}\rangle_{\rm lc} > \langle\phi_{\rm lc}\rangle$ and $\langle\phi_{\rm gb}\rangle_{\rm lc} \simeq \langle\phi_{\rm gb}\rangle$. In this state, particle sizes cycle periodically in a locally coordinated way, yielding the spontaneous emergence of deformation waves [Figs.~\ref{fig:phase_diagram_local}(d) and~\ref{fig:phase_diagram_local}(e)] not present in the unbiased dynamics [Fig.~\ref{fig:phase_diagram_local}(c)]. For higher $|s|$, the range of particle coordination increases, which increases global order ($\langle\phi_{\rm gb}\rangle_{\rm lc} > \langle\phi_{\rm gb}\rangle$) and ultimately results in a cycling state [Fig.~\ref{fig:phase_diagram_local}(f)] similar to that of the previous BE [Fig.~\ref{fig:N32_trajs}(e)].

In this manner, the BE promoting local order [Eq.~\eqref{eq:slc}] reproduces all the states of synchronizing PAM~\cite{PulsatingAM, Manacorda2023}: disorder, arrest, cycles, and waves. Furthermore, waves only arise for box sizes $L$ accommodating at least one wavelength. As the wavelength increases with $|s|$, waves are only stable over a finite range of $s$. As such, waves can be seen as precursory to cycles. In contrast, arrest does not display such a gradual ordering from local to global, but rather directly emerges from disorder at comparatively low bias. Moreover, the critical $s$ for this transition is almost unchanged when biasing with either global [Fig.~\ref{fig:phase_diagram}(a)] or local [Fig.~\ref{fig:phase_diagram_local}(b)] order.

In short, our results demonstrate that waves spontaneously emerge as a strategy to promote high {\em local} order, while maintaining only moderate {\em global} order. Again, the box geometry plays a crucial role here. Specifically, for values of $\Lol$ promoting regular packing configurations, local deformations are strongly hampered, so that arrest is more stable than waves for any $s<0$.


\textit{Discussion.}---We reveal some unexpected connections between packing configurations and rare fluctuations in dense systems of pulsating particles. The box geometry is a proxy to controlling the packing structure, with dramatic consequences on collective effects. Specifically, we show that one can induce transitions between two types of ordered states in BEs, either with or without current, simply by changing the box geometry. Arrest is associated with regular packing configurations where the particle repulsion exactly counteracts their pulsation. On the other hand, when the box geometry induces a defective packing, it generates regions of inhomogeneous repulsion that ultimately lead to cycling. Remarkably, the transitions between arrest and cycles in the biased dynamics can actually be anticipated from the statistics of current and order in the unbiased dynamics.

Our transitions bring interesting parallels with the emergence of arrest in other dense systems. In densely packed SPPs, structural defects destabilize arrest~\cite{SPPHexFluidization}, alter the glass transition~\cite{glassyAM}, and induce intermittent plastic yielding~\cite{PlasticAM}. The mechanical properties of such systems can actually be related to those of sheared granular systems~\cite{Manning2021}. Moreover, local growth of deforming particles also results in dynamical heterogeneities resembling sheared glasses~\cite{GrowingAMGlassAnalogies}. These examples suggest that a generic mechanism may explain how activity controls the transitions between arrested and fluidized states. Remarkably, even in the absence of shear, allowing size fluctuations shifts the glass transition to lower temperatures~\cite{Berthier2017, SWAPglass, particle_size_change_jamming, ideal_glass_radii_change_method}, illustrating how local deformation helps relax the dynamics near arrest~\cite{Manning2023}.

Our approach could also motivate further studies in other active models where synchronization yields patterns~\cite{Kuramoto1986, Strogatz2017, Reichhardt2023}. For instance, considering BEs with local or global order could help delineate minimal conditions to stabilize patterns, similarly to how waves only emerge for specific box geometries in our case. To improve sampling, one could rely on more complex interactions beyond the synchronization considered here. To this end, recent methods inspired by machine learning provide a rich toolbox~\cite{Whitelam2020, Garrahan2021, Rotskoff2022} which could prove quite useful.
\newline



\acknowledgments{Supporting code and data is freely available, respectively, in 
Github: \href{https://github.com/CreditDefaultSwap/pulsating_active_matter_popdyn}{CreditDefaultSwap/pulsating\_active\_matter\_popdyn},
and in the Zenodo data repository: 
\href{https://doi.org/10.5281/zenodo.14588892}{10.5281/zenodo.14588892}.
We acknowledge insightful discussions with Y.-E. Keta and R. L. Jack. Work funded by the Luxembourg National Research Fund (FNR), grant reference 14389168, and the Marie Skłodowska-Curie grant No. 101104945.}
\newline


\textit{Appendix A: Overlap distance in dense monodisperse systems.}--- 
To compute the overlap distance $\sigma_{\rm ovx}$ for a monodisperse crystal, we determine the maximum interparticle distance $r_m$ which minimizes repulsion between particles by minimizing their overlap. Monodispersity implies that all particles have the same phase $\theta_i=\theta$. For arrested states ($\dot\theta=0$), the noiseless phase dynamics [Eq.~\eqref{eq:dynamics}] reduces to
\begin{equation}
    \omega = n_n\mu_\theta \partial{_{\theta}} U(r,\theta) ,
\end{equation}
where $n_n$ is the number of contact neighbors, which defines an implicit function $r=r(\theta)$. Assuming hexagonal packing ($n_n=6$), and maximizing with respect to $\theta$, we find $\theta_m \simeq0.562$, and deduce $r(\theta_m)\simeq0.845$. The overlap distance follows as $\sigma_{\rm ovx} = 2 \sigma_m - r_m \simeq 0.133$, where $\sigma_m = [1+\lambda\sin(\theta_m)]/[2(1+\lambda)]$ is the optimal particle radius [Eq.~\eqref{eq:sigma}].


\begin{figure*}
  \centering
  \includegraphics[scale=.3]{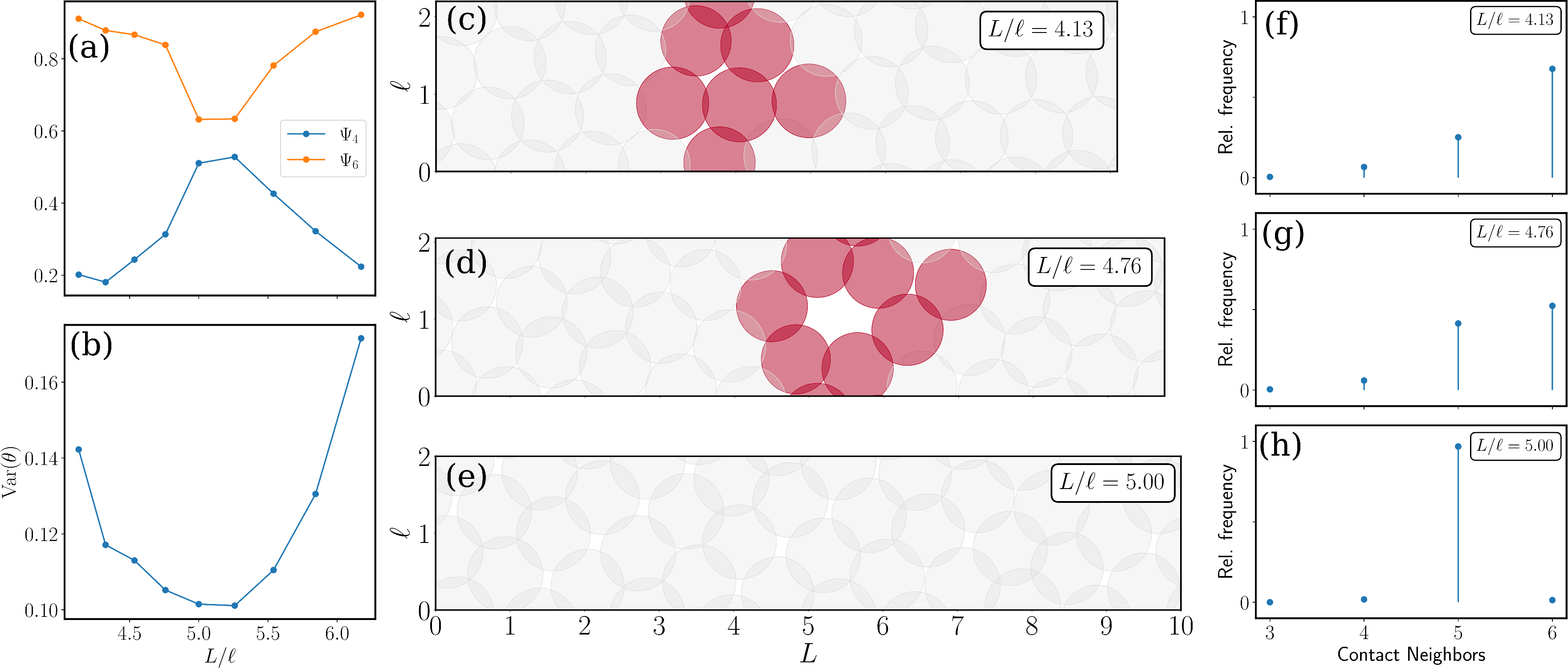}
  \caption{(a)~Orientational order parameters $\Psi_{4/6}$ and (b)~phase variance $\Var(\theta)$ in ensembles biased by global order [Eq.~\eqref{eq:sgb}], as functions of the box geometry $\Lol$ for $s=-6$. (c-e)~Corresponding snapshots of biased configurations, identical to Fig.~\ref{fig:N32_trajs}, here colored by the number of contact neighbors. For (c-d), red indicates particles with 5 contact neighbors or less, and gray for 6 neighbors. For (e), gray indicates particles in contact with 5 neighbors. (f-h)~Histogram in time of the number of contacts estimated from biased trajectories, with same parameters as in panels (c-e) .}
\label{fig:structure_analysis}
\end{figure*}

\textit{Appendix B: Structural analysis of biased configurations.}---To systematically characterize the structural configurations emerging under bias, we measure the orientational order parameter $\Psi_k$ defined as
\begin{equation}\label{eq:psi}
    \Psi_k = \frac{1}{N} \Bigg\langle \sum_{j=1}^N \sum_{l=1}^{n_j} e^{i k \alpha_{jl}} \Bigg\rangle_{\rm gb} ,
\end{equation}
where $n_j$ is the number of nearest neighbors of particle $j$, and the integer $k$ refers to the degree of orientational order. The orientation angle $\alpha_{jl}$ is defined in terms of the distance vector between two neighboring particles $(j,k)$ relative to a fixed axis. We choose to consider $k=\{4,6\}$, and we employ the weighed Voronoi method to evaluate $\Psi_{4/6}~$\cite{weighed_voronoi, freud_package}. The system goes from hexagonal-like to square-like structures as $\Lol$ varies from arrest ($\Lol\simeq5$, high $\Psi_4$, and low $\Psi_6$) to cycling and intermittent states ($\Lol\neq 5$, high $\Psi_6$, and low $\Psi_4$)[Fig.~\ref{fig:structure_analysis}(a)].

Interactions between particle phases $\theta_i$ take place whenever particles are in contact ($a_{ij} \le 1$), so that structural order can only emerge via the following two conditions: (i)~low particle polydispersity (i.e., phase homogeneity), and (ii)~persistent neighborhood of particle contacts. The latter accounts for the fact that irregular contact profiles (e.g., via structural defects) alter the local phase dynamics and promote polydispersity. In contrast, for passive particles with fixed sizes, interparticle distance alone suffices to determine structural order.

To evaluate the statistics of particle polydispersity, we first measure the phase variance 
\begin{equation}\label{eq:var}
    \Var(\theta) = \frac{1}{N} \Bigg\langle \sum_{i=1}^N (\theta_i-\theta_m)^2 \Bigg\rangle_{\rm gb}
\end{equation}  
for realizations where the average phase $\theta_m=\frac 1 N \sum_{j=1}^N \theta_j$ is approximately the same for all states (either arrested, intermittent, or cycling). We find that $\Var(\theta)$ is lowest for arrest ($\Lol=5$), and higher for the intermittent and cycling states ($\Lol=4.13$ and $6.17$) [Fig.~\ref{fig:structure_analysis}(b)]. This result indicates that arrest features a more homogeneous phase distribution (i.e., lower polydispersity) than intermittent or cycling states.

Second, we display the contact profiles, given by the number of contact neighbors per particle [Figs.~\ref{fig:structure_analysis}(c-e)], and estimate their persistence in time via histograms of the number of contacts from the corresponding biased trajectories [Figs.~\ref{fig:structure_analysis}(f-h)]. For intermittent and cycling states, histograms reveal a majority of six-neighbor contacts, along with noticeable contributions from contacts with five or fewer neighbors [Figs.~\ref{fig:structure_analysis}(f-g)]: this result hints at the presence of defects in a hexagonal packing structure. In contrast, five-neighbor contacts predominate for  arrest [Fig.~\ref{fig:structure_analysis}(h)], indicating a persistent, regular structure, albeit with a nonhexagonal packing. Indeed, the arrested structure conforms to an elongated honeycomb crystal, as shown in Fig.~S4 of~\cite{SM}.

In all, the structural analysis supports that arrest is associated with a regular packing configuration, whereas intermittent and cycling states feature some defective and non-persistent configurations.



\begin{thebibliography}{67}%
\makeatletter
\providecommand \@ifxundefined [1]{%
 \@ifx{#1\undefined}
}%
\providecommand \@ifnum [1]{%
 \ifnum #1\expandafter \@firstoftwo
 \else \expandafter \@secondoftwo
 \fi
}%
\providecommand \@ifx [1]{%
 \ifx #1\expandafter \@firstoftwo
 \else \expandafter \@secondoftwo
 \fi
}%
\providecommand \natexlab [1]{#1}%
\providecommand \enquote  [1]{``#1''}%
\providecommand \bibnamefont  [1]{#1}%
\providecommand \bibfnamefont [1]{#1}%
\providecommand \citenamefont [1]{#1}%
\providecommand \href@noop [0]{\@secondoftwo}%
\providecommand \href [0]{\begingroup \@sanitize@url \@href}%
\providecommand \@href[1]{\@@startlink{#1}\@@href}%
\providecommand \@@href[1]{\endgroup#1\@@endlink}%
\providecommand \@sanitize@url [0]{\catcode `\\12\catcode `\$12\catcode
  `\&12\catcode `\#12\catcode `\^12\catcode `\_12\catcode `\%12\relax}%
\providecommand \@@startlink[1]{}%
\providecommand \@@endlink[0]{}%
\providecommand \url  [0]{\begingroup\@sanitize@url \@url }%
\providecommand \@url [1]{\endgroup\@href {#1}{\urlprefix }}%
\providecommand \urlprefix  [0]{URL }%
\providecommand \Eprint [0]{\href }%
\providecommand \doibase [0]{https://doi.org/}%
\providecommand \selectlanguage [0]{\@gobble}%
\providecommand \bibinfo  [0]{\@secondoftwo}%
\providecommand \bibfield  [0]{\@secondoftwo}%
\providecommand \translation [1]{[#1]}%
\providecommand \BibitemOpen [0]{}%
\providecommand \bibitemStop [0]{}%
\providecommand \bibitemNoStop [0]{.\EOS\space}%
\providecommand \EOS [0]{\spacefactor3000\relax}%
\providecommand \BibitemShut  [1]{\csname bibitem#1\endcsname}%
\let\auto@bib@innerbib\@empty
\bibitem [{\citenamefont {Marchetti}\ \emph {et~al.}(2013)\citenamefont
  {Marchetti}, \citenamefont {Joanny}, \citenamefont {Ramaswamy}, \citenamefont
  {Liverpool}, \citenamefont {Prost}, \citenamefont {Rao},\ and\ \citenamefont
  {Simha}}]{Marchetti2013}%
  \BibitemOpen
  \bibfield  {author} {\bibinfo {author} {\bibfnamefont {M.~C.}\ \bibnamefont
  {Marchetti}}, \bibinfo {author} {\bibfnamefont {J.~F.}\ \bibnamefont
  {Joanny}}, \bibinfo {author} {\bibfnamefont {S.}~\bibnamefont {Ramaswamy}},
  \bibinfo {author} {\bibfnamefont {T.~B.}\ \bibnamefont {Liverpool}}, \bibinfo
  {author} {\bibfnamefont {J.}~\bibnamefont {Prost}}, \bibinfo {author}
  {\bibfnamefont {M.}~\bibnamefont {Rao}},\ and\ \bibinfo {author}
  {\bibfnamefont {R.~A.}\ \bibnamefont {Simha}},\ }\bibfield  {title} {\bibinfo
  {title} {Hydrodynamics of soft active matter},\ }\href
  {https://doi.org/10.1103/RevModPhys.85.1143} {\bibfield  {journal} {\bibinfo
  {journal} {Rev. Mod. Phys.}\ }\textbf {\bibinfo {volume} {85}},\ \bibinfo
  {pages} {1143} (\bibinfo {year} {2013})}\BibitemShut {NoStop}%
\bibitem [{\citenamefont {Bechinger}\ \emph {et~al.}(2016)\citenamefont
  {Bechinger}, \citenamefont {Di~Leonardo}, \citenamefont {L\"owen},
  \citenamefont {Reichhardt}, \citenamefont {Volpe},\ and\ \citenamefont
  {Volpe}}]{Bechinger2016}%
  \BibitemOpen
  \bibfield  {author} {\bibinfo {author} {\bibfnamefont {C.}~\bibnamefont
  {Bechinger}}, \bibinfo {author} {\bibfnamefont {R.}~\bibnamefont
  {Di~Leonardo}}, \bibinfo {author} {\bibfnamefont {H.}~\bibnamefont
  {L\"owen}}, \bibinfo {author} {\bibfnamefont {C.}~\bibnamefont {Reichhardt}},
  \bibinfo {author} {\bibfnamefont {G.}~\bibnamefont {Volpe}},\ and\ \bibinfo
  {author} {\bibfnamefont {G.}~\bibnamefont {Volpe}},\ }\bibfield  {title}
  {\bibinfo {title} {Active particles in complex and crowded environments},\
  }\href {https://doi.org/10.1103/RevModPhys.88.045006} {\bibfield  {journal}
  {\bibinfo  {journal} {Rev. Mod. Phys.}\ }\textbf {\bibinfo {volume} {88}},\
  \bibinfo {pages} {045006} (\bibinfo {year} {2016})}\BibitemShut {NoStop}%
\bibitem [{\citenamefont {Fodor}\ and\ \citenamefont
  {C. Marchetti}(2018)}]{Marchetti2018}%
  \BibitemOpen
  \bibfield  {author} {\bibinfo {author} {\bibfnamefont {{\'E}.}~\bibnamefont
  {Fodor}}\ and\ \bibinfo {author} {\bibfnamefont {M.}~\bibnamefont
  {C. Marchetti}},\ }\bibfield  {title} {\bibinfo {title} {The statistical
  physics of active matter: From self-catalytic colloids to living cells},\
  }\href {https://doi.org/10.1016/j.physa.2017.12.137} {\bibfield  {journal}
  {\bibinfo  {journal} {Physica A}\ }\textbf {\bibinfo {volume} {504}},\
  \bibinfo {pages} {106} (\bibinfo {year} {2018})}\BibitemShut {NoStop}%
\bibitem [{\citenamefont {Vicsek}\ \emph {et~al.}(1995)\citenamefont {Vicsek},
  \citenamefont {Czir\'ok}, \citenamefont {Ben-Jacob}, \citenamefont {Cohen},\
  and\ \citenamefont {Shochet}}]{Vicsek1995}%
  \BibitemOpen
  \bibfield  {author} {\bibinfo {author} {\bibfnamefont {T.}~\bibnamefont
  {Vicsek}}, \bibinfo {author} {\bibfnamefont {A.}~\bibnamefont {Czir\'ok}},
  \bibinfo {author} {\bibfnamefont {E.}~\bibnamefont {Ben-Jacob}}, \bibinfo
  {author} {\bibfnamefont {I.}~\bibnamefont {Cohen}},\ and\ \bibinfo {author}
  {\bibfnamefont {O.}~\bibnamefont {Shochet}},\ }\bibfield  {title} {\bibinfo
  {title} {Novel type of phase transition in a system of self-driven
  particles},\ }\href {https://doi.org/10.1103/PhysRevLett.75.1226} {\bibfield
  {journal} {\bibinfo  {journal} {Phys. Rev. Lett.}\ }\textbf {\bibinfo
  {volume} {75}},\ \bibinfo {pages} {1226} (\bibinfo {year}
  {1995})}\BibitemShut {NoStop}%
\bibitem [{\citenamefont {Chat\'e}(2020)}]{Chate2020}%
  \BibitemOpen
  \bibfield  {author} {\bibinfo {author} {\bibfnamefont {H.}~\bibnamefont
  {Chat\'e}},\ }\bibfield  {title} {\bibinfo {title} {Dry aligning dilute
  active matter},\ }\href
  {https://doi.org/10.1146/annurev-conmatphys-031119-050752} {\bibfield
  {journal} {\bibinfo  {journal} {Annu. Rev. Condens. Matter Phys.}\ }\textbf
  {\bibinfo {volume} {11}},\ \bibinfo {pages} {189} (\bibinfo {year}
  {2020})}\BibitemShut {NoStop}%
\bibitem [{\citenamefont {Fily}\ and\ \citenamefont {Marchetti}(2012)}]{MIPS1}%
  \BibitemOpen
  \bibfield  {author} {\bibinfo {author} {\bibfnamefont {Y.}~\bibnamefont
  {Fily}}\ and\ \bibinfo {author} {\bibfnamefont {M.~C.}\ \bibnamefont
  {Marchetti}},\ }\bibfield  {title} {\bibinfo {title} {Athermal phase
  separation of self-propelled particles with no alignment},\ }\href
  {https://doi.org/10.1103/PhysRevLett.108.235702} {\bibfield  {journal}
  {\bibinfo  {journal} {Phys. Rev. Lett.}\ }\textbf {\bibinfo {volume} {108}},\
  \bibinfo {pages} {235702} (\bibinfo {year} {2012})}\BibitemShut {NoStop}%
\bibitem [{\citenamefont {Cates}\ and\ \citenamefont
  {Tailleur}(2015)}]{MIPSrev}%
  \BibitemOpen
  \bibfield  {author} {\bibinfo {author} {\bibfnamefont {M.~E.}\ \bibnamefont
  {Cates}}\ and\ \bibinfo {author} {\bibfnamefont {J.}~\bibnamefont
  {Tailleur}},\ }\bibfield  {title} {\bibinfo {title} {Motility-induced phase
  separation},\ }\href
  {https://doi.org/10.1146/annurev-conmatphys-031214-014710} {\bibfield
  {journal} {\bibinfo  {journal} {Annu. Rev. Condens. Matter Phys.}\ }\textbf
  {\bibinfo {volume} {6}},\ \bibinfo {pages} {219} (\bibinfo {year}
  {2015})}\BibitemShut {NoStop}%
\bibitem [{\citenamefont {Bailles}\ \emph {et~al.}(2022)\citenamefont
  {Bailles}, \citenamefont {Gehrels},\ and\ \citenamefont
  {Lecuit}}]{Lecuit2022}%
  \BibitemOpen
  \bibfield  {author} {\bibinfo {author} {\bibfnamefont {A.}~\bibnamefont
  {Bailles}}, \bibinfo {author} {\bibfnamefont {E.~W.}\ \bibnamefont
  {Gehrels}},\ and\ \bibinfo {author} {\bibfnamefont {T.}~\bibnamefont
  {Lecuit}},\ }\bibfield  {title} {\bibinfo {title} {Mechanochemical principles
  of spatial and temporal patterns in cells and tissues},\ }\href
  {https://doi.org/10.1146/annurev-cellbio-120420-095337} {\bibfield  {journal}
  {\bibinfo  {journal} {Annu. Rev. Cell Dev. Biol.}\ }\textbf {\bibinfo
  {volume} {38}},\ \bibinfo {pages} {321} (\bibinfo {year} {2022})}\BibitemShut
  {NoStop}%
\bibitem [{\citenamefont {Karma}(2013)}]{Karma2013}%
  \BibitemOpen
  \bibfield  {author} {\bibinfo {author} {\bibfnamefont {A.}~\bibnamefont
  {Karma}},\ }\bibfield  {title} {\bibinfo {title} {Physics of cardiac
  arrhythmogenesis},\ }\href
  {https://doi.org/10.1146/annurev-conmatphys-020911-125112} {\bibfield
  {journal} {\bibinfo  {journal} {Annu. Rev. Condens. Matter Phys.}\ }\textbf
  {\bibinfo {volume} {4}},\ \bibinfo {pages} {313} (\bibinfo {year}
  {2013})}\BibitemShut {NoStop}%
\bibitem [{\citenamefont {Myers}\ and\ \citenamefont {Elad}(2017)}]{Elad2017}%
  \BibitemOpen
  \bibfield  {author} {\bibinfo {author} {\bibfnamefont {K.~M.}\ \bibnamefont
  {Myers}}\ and\ \bibinfo {author} {\bibfnamefont {D.}~\bibnamefont {Elad}},\
  }\bibfield  {title} {\bibinfo {title} {Biomechanics of the human uterus},\
  }\href {https://doi.org/https://doi.org/10.1002/wsbm.1388} {\bibfield
  {journal} {\bibinfo  {journal} {WIREs Syst. Biol. Med.}\ }\textbf {\bibinfo
  {volume} {9}},\ \bibinfo {pages} {e1388} (\bibinfo {year}
  {2017})}\BibitemShut {NoStop}%
\bibitem [{\citenamefont {Zhang}\ and\ \citenamefont
  {Fodor}(2023)}]{PulsatingAM}%
  \BibitemOpen
  \bibfield  {author} {\bibinfo {author} {\bibfnamefont {Y.}~\bibnamefont
  {Zhang}}\ and\ \bibinfo {author} {\bibfnamefont {E.}~\bibnamefont {Fodor}},\
  }\bibfield  {title} {\bibinfo {title} {Pulsating active matter},\ }\href
  {https://doi.org/10.1103/PhysRevLett.131.238302} {\bibfield  {journal}
  {\bibinfo  {journal} {Phys. Rev. Lett.}\ }\textbf {\bibinfo {volume} {131}},\
  \bibinfo {pages} {238302} (\bibinfo {year} {2023})}\BibitemShut {NoStop}%
\bibitem [{\citenamefont {Tjhung}\ and\ \citenamefont
  {Kawasaki}(2017)}]{Tjhung2016}%
  \BibitemOpen
  \bibfield  {author} {\bibinfo {author} {\bibfnamefont {E.}~\bibnamefont
  {Tjhung}}\ and\ \bibinfo {author} {\bibfnamefont {T.}~\bibnamefont
  {Kawasaki}},\ }\bibfield  {title} {\bibinfo {title} {Excitation of
  vibrational soft modes in disordered systems using active oscillation},\
  }\href {https://doi.org/10.1039/C6SM00788K} {\bibfield  {journal} {\bibinfo
  {journal} {{Soft Matter}}\ }\textbf {\bibinfo {volume} {13}},\ \bibinfo
  {pages} {111} (\bibinfo {year} {2017})}\BibitemShut {NoStop}%
\bibitem [{\citenamefont {Tjhung}\ and\ \citenamefont
  {Berthier}(2017)}]{Tjhung2017}%
  \BibitemOpen
  \bibfield  {author} {\bibinfo {author} {\bibfnamefont {E.}~\bibnamefont
  {Tjhung}}\ and\ \bibinfo {author} {\bibfnamefont {L.}~\bibnamefont
  {Berthier}},\ }\bibfield  {title} {\bibinfo {title} {Discontinuous zation
  transition in time-correlated assemblies of actively deforming particles},\
  }\href {https://doi.org/10.1103/PhysRevE.96.050601} {\bibfield  {journal}
  {\bibinfo  {journal} {Phys. Rev. E}\ }\textbf {\bibinfo {volume} {96}},\
  \bibinfo {pages} {050601} (\bibinfo {year} {2017})}\BibitemShut {NoStop}%
\bibitem [{\citenamefont {Oyama}\ \emph {et~al.}(2019)\citenamefont {Oyama},
  \citenamefont {Kawasaki}, \citenamefont {Mizuno},\ and\ \citenamefont
  {Ikeda}}]{Oyama2019}%
  \BibitemOpen
  \bibfield  {author} {\bibinfo {author} {\bibfnamefont {N.}~\bibnamefont
  {Oyama}}, \bibinfo {author} {\bibfnamefont {T.}~\bibnamefont {Kawasaki}},
  \bibinfo {author} {\bibfnamefont {H.}~\bibnamefont {Mizuno}},\ and\ \bibinfo
  {author} {\bibfnamefont {A.}~\bibnamefont {Ikeda}},\ }\bibfield  {title}
  {\bibinfo {title} {Glassy dynamics of a model of bacterial cytoplasm with
  metabolic activities},\ }\href
  {https://doi.org/10.1103/PhysRevResearch.1.032038} {\bibfield  {journal}
  {\bibinfo  {journal} {Phys. Rev. Research}\ }\textbf {\bibinfo {volume}
  {1}},\ \bibinfo {pages} {032038} (\bibinfo {year} {2019})}\BibitemShut
  {NoStop}%
\bibitem [{\citenamefont {Koyano}\ \emph {et~al.}(2019)\citenamefont {Koyano},
  \citenamefont {Kitahata},\ and\ \citenamefont {Mikhailov}}]{Koyano2019}%
  \BibitemOpen
  \bibfield  {author} {\bibinfo {author} {\bibfnamefont {Y.}~\bibnamefont
  {Koyano}}, \bibinfo {author} {\bibfnamefont {H.}~\bibnamefont {Kitahata}},\
  and\ \bibinfo {author} {\bibfnamefont {A.~S.}\ \bibnamefont {Mikhailov}},\
  }\bibfield  {title} {\bibinfo {title} {Diffusion in crowded colloids of
  particles cyclically changing their shapes},\ }\href
  {https://doi.org/10.1209/0295-5075/128/40003} {\bibfield  {journal} {\bibinfo
   {journal} {EPL}\ }\textbf {\bibinfo {volume} {128}},\ \bibinfo {pages}
  {40003} (\bibinfo {year} {2019})}\BibitemShut {NoStop}%
\bibitem [{\citenamefont {Togashi}(2019)}]{Togashi2019}%
  \BibitemOpen
  \bibfield  {author} {\bibinfo {author} {\bibfnamefont {Y.}~\bibnamefont
  {Togashi}},\ }\bibfield  {title} {\bibinfo {title} {{Modeling of
  nanomachine/micromachine crowds: Interplay between the internal state and
  surroundings}},\ }\href {https://doi.org/10.1021/acs.jpcb.8b10633} {\bibfield
   {journal} {\bibinfo  {journal} {J. Phys. Chem. B}\ }\textbf {\bibinfo
  {volume} {123}},\ \bibinfo {pages} {1481} (\bibinfo {year}
  {2019})}\BibitemShut {NoStop}%
\bibitem [{\citenamefont {Manacorda}\ and\ \citenamefont {Étienne
  Fodor}(2024)}]{Manacorda2024}%
  \BibitemOpen
  \bibfield  {author} {\bibinfo {author} {\bibfnamefont {A.}~\bibnamefont
  {Manacorda}}\ and\ \bibinfo {author} {\bibnamefont {Étienne Fodor}},\
  }\href@noop {} {\bibinfo {title} {Diffusive oscillators capture the pulsating
  states of deformable particles}} (\bibinfo {year} {2024}),\ \Eprint
  {https://arxiv.org/abs/2310.14370} {arXiv:2310.14370 [cond-mat.stat-mech]}
  \BibitemShut {NoStop}%
\bibitem [{\citenamefont {hua Liu}\ \emph {et~al.}(2024)\citenamefont {hua
  Liu}, \citenamefont {jing Zhu},\ and\ \citenamefont {quan Ai}}]{Liu2024}%
  \BibitemOpen
  \bibfield  {author} {\bibinfo {author} {\bibfnamefont {W.}~\bibnamefont {hua
  Liu}}, \bibinfo {author} {\bibfnamefont {W.}~\bibnamefont {jing Zhu}},\ and\
  \bibinfo {author} {\bibfnamefont {B.}~\bibnamefont {quan Ai}},\ }\bibfield
  {title} {\bibinfo {title} {Collective motion of pulsating active particles in
  confined structures},\ }\href {https://doi.org/10.1088/1367-2630/ad23a5}
  {\bibfield  {journal} {\bibinfo  {journal} {New J. Phys.}\ }\textbf {\bibinfo
  {volume} {26}},\ \bibinfo {pages} {023017} (\bibinfo {year}
  {2024})}\BibitemShut {NoStop}%
\bibitem [{\citenamefont {Li}\ \emph {et~al.}(2024)\citenamefont {Li},
  \citenamefont {Lei},\ and\ \citenamefont {qiang Ma}}]{Li2024}%
  \BibitemOpen
  \bibfield  {author} {\bibinfo {author} {\bibfnamefont {Z.-Q.}\ \bibnamefont
  {Li}}, \bibinfo {author} {\bibfnamefont {Q.-L.}\ \bibnamefont {Lei}},\ and\
  \bibinfo {author} {\bibfnamefont {Y.}~\bibnamefont {qiang Ma}},\ }\bibfield
  {title} {\bibinfo {title} {Fluidization and anomalous density fluctuations in
  epithelial tissues with pulsating activity},\ }\href@noop {} {\bibfield
  {journal} {\bibinfo  {journal} {ArXiv e-prints}\ } (\bibinfo {year}
  {2024})},\ \Eprint {https://arxiv.org/abs/2402.02981} {arXiv:2402.02981}
  \BibitemShut {NoStop}%
\bibitem [{\citenamefont {Jack}(2020)}]{Jack2020}%
  \BibitemOpen
  \bibfield  {author} {\bibinfo {author} {\bibfnamefont {R.~L.}\ \bibnamefont
  {Jack}},\ }\bibfield  {title} {\bibinfo {title} {Ergodicity and large
  deviations in physical systems with stochastic dynamics},\ }\href
  {https://doi.org/10.1140/epjb/e2020-100605-3} {\bibfield  {journal} {\bibinfo
   {journal} {Eur. Phys. J. B}\ }\textbf {\bibinfo {volume} {93}},\ \bibinfo
  {pages} {74} (\bibinfo {year} {2020})}\BibitemShut {NoStop}%
\bibitem [{\citenamefont {Touchette}(2009)}]{Touchette2009}%
  \BibitemOpen
  \bibfield  {author} {\bibinfo {author} {\bibfnamefont {H.}~\bibnamefont
  {Touchette}},\ }\bibfield  {title} {\bibinfo {title} {The large deviation
  approach to statistical mechanics},\ }\href
  {https://doi.org/10.1016/j.physrep.2009.05.002} {\bibfield  {journal}
  {\bibinfo  {journal} {Phys. Rep.}\ }\textbf {\bibinfo {volume} {478}},\
  \bibinfo {pages} {1} (\bibinfo {year} {2009})}\BibitemShut {NoStop}%
\bibitem [{\citenamefont {Garrahan}\ \emph {et~al.}(2007)\citenamefont
  {Garrahan}, \citenamefont {Jack}, \citenamefont {Lecomte}, \citenamefont
  {Pitard}, \citenamefont {van Duijvendijk},\ and\ \citenamefont {van
  Wijland}}]{Wijland2007}%
  \BibitemOpen
  \bibfield  {author} {\bibinfo {author} {\bibfnamefont {J.~P.}\ \bibnamefont
  {Garrahan}}, \bibinfo {author} {\bibfnamefont {R.~L.}\ \bibnamefont {Jack}},
  \bibinfo {author} {\bibfnamefont {V.}~\bibnamefont {Lecomte}}, \bibinfo
  {author} {\bibfnamefont {E.}~\bibnamefont {Pitard}}, \bibinfo {author}
  {\bibfnamefont {K.}~\bibnamefont {van Duijvendijk}},\ and\ \bibinfo {author}
  {\bibfnamefont {F.}~\bibnamefont {van Wijland}},\ }\bibfield  {title}
  {\bibinfo {title} {Dynamical first-order phase transition in kinetically
  constrained models of glasses},\ }\href
  {https://doi.org/10.1103/PhysRevLett.98.195702} {\bibfield  {journal}
  {\bibinfo  {journal} {Phys. Rev. Lett.}\ }\textbf {\bibinfo {volume} {98}},\
  \bibinfo {pages} {195702} (\bibinfo {year} {2007})}\BibitemShut {NoStop}%
\bibitem [{\citenamefont {Tiz\'on-Escamilla}\ \emph {et~al.}(2017)\citenamefont
  {Tiz\'on-Escamilla}, \citenamefont {P\'erez-Espigares}, \citenamefont
  {Garrido},\ and\ \citenamefont {Hurtado}}]{Hurtado2017}%
  \BibitemOpen
  \bibfield  {author} {\bibinfo {author} {\bibfnamefont {N.}~\bibnamefont
  {Tiz\'on-Escamilla}}, \bibinfo {author} {\bibfnamefont {C.}~\bibnamefont
  {P\'erez-Espigares}}, \bibinfo {author} {\bibfnamefont {P.~L.}\ \bibnamefont
  {Garrido}},\ and\ \bibinfo {author} {\bibfnamefont {P.~I.}\ \bibnamefont
  {Hurtado}},\ }\bibfield  {title} {\bibinfo {title} {Order and symmetry
  breaking in the fluctuations of driven systems},\ }\href
  {https://doi.org/10.1103/PhysRevLett.119.090602} {\bibfield  {journal}
  {\bibinfo  {journal} {Phys. Rev. Lett.}\ }\textbf {\bibinfo {volume} {119}},\
  \bibinfo {pages} {090602} (\bibinfo {year} {2017})}\BibitemShut {NoStop}%
\bibitem [{\citenamefont {Royall}\ \emph {et~al.}(2020)\citenamefont {Royall},
  \citenamefont {Turci},\ and\ \citenamefont {Speck}}]{glassStructureLD}%
  \BibitemOpen
  \bibfield  {author} {\bibinfo {author} {\bibfnamefont {C.~P.}\ \bibnamefont
  {Royall}}, \bibinfo {author} {\bibfnamefont {F.}~\bibnamefont {Turci}},\ and\
  \bibinfo {author} {\bibfnamefont {T.}~\bibnamefont {Speck}},\ }\bibfield
  {title} {\bibinfo {title} {{Dynamical phase transitions and their relation to
  structural and thermodynamic aspects of glass physics}},\ }\href
  {https://doi.org/10.1063/5.0006998} {\bibfield  {journal} {\bibinfo
  {journal} {J. Chem. Phys.}\ }\textbf {\bibinfo {volume} {153}},\ \bibinfo
  {pages} {090901} (\bibinfo {year} {2020})}\BibitemShut {NoStop}%
\bibitem [{\citenamefont {Chetrite}\ and\ \citenamefont
  {Touchette}(2015)}]{Chetrite2015}%
  \BibitemOpen
  \bibfield  {author} {\bibinfo {author} {\bibfnamefont {R.}~\bibnamefont
  {Chetrite}}\ and\ \bibinfo {author} {\bibfnamefont {H.}~\bibnamefont
  {Touchette}},\ }\bibfield  {title} {\bibinfo {title} {Nonequilibrium markov
  processes conditioned on large deviations},\ }\href
  {https://doi.org/10.1007/s00023-014-0375-8} {\bibfield  {journal} {\bibinfo
  {journal} {Annales Henri Poincar{\'e}}\ }\textbf {\bibinfo {volume} {16}},\
  \bibinfo {pages} {2005} (\bibinfo {year} {2015})}\BibitemShut {NoStop}%
\bibitem [{\citenamefont {Jack}\ and\ \citenamefont
  {Sollich}(2015)}]{Jack2015}%
  \BibitemOpen
  \bibfield  {author} {\bibinfo {author} {\bibfnamefont {R.~L.}\ \bibnamefont
  {Jack}}\ and\ \bibinfo {author} {\bibfnamefont {P.}~\bibnamefont {Sollich}},\
  }\bibfield  {title} {\bibinfo {title} {Effective interactions and large
  deviations in stochastic processes},\ }\href
  {https://doi.org/10.1140/epjst/e2015-02416-9} {\bibfield  {journal} {\bibinfo
   {journal} {Eur. Phys. J. Special Topics}\ }\textbf {\bibinfo {volume}
  {224}},\ \bibinfo {pages} {2351} (\bibinfo {year} {2015})}\BibitemShut
  {NoStop}%
\bibitem [{\citenamefont {Ray}\ \emph {et~al.}(2018)\citenamefont {Ray},
  \citenamefont {Chan},\ and\ \citenamefont {Limmer}}]{Limmer2018}%
  \BibitemOpen
  \bibfield  {author} {\bibinfo {author} {\bibfnamefont {U.}~\bibnamefont
  {Ray}}, \bibinfo {author} {\bibfnamefont {G.~K.-L.}\ \bibnamefont {Chan}},\
  and\ \bibinfo {author} {\bibfnamefont {D.~T.}\ \bibnamefont {Limmer}},\
  }\bibfield  {title} {\bibinfo {title} {Exact fluctuations of nonequilibrium
  steady states from approximate auxiliary dynamics},\ }\href
  {https://doi.org/10.1103/PhysRevLett.120.210602} {\bibfield  {journal}
  {\bibinfo  {journal} {Phys. Rev. Lett.}\ }\textbf {\bibinfo {volume} {120}},\
  \bibinfo {pages} {210602} (\bibinfo {year} {2018})}\BibitemShut {NoStop}%
\bibitem [{\citenamefont {Tociu}\ \emph {et~al.}(2019)\citenamefont {Tociu},
  \citenamefont {Fodor}, \citenamefont {Nemoto},\ and\ \citenamefont
  {Vaikuntanathan}}]{Suri2019}%
  \BibitemOpen
  \bibfield  {author} {\bibinfo {author} {\bibfnamefont {L.}~\bibnamefont
  {Tociu}}, \bibinfo {author} {\bibfnamefont {{\'E}.}~\bibnamefont {Fodor}},
  \bibinfo {author} {\bibfnamefont {T.}~\bibnamefont {Nemoto}},\ and\ \bibinfo
  {author} {\bibfnamefont {S.}~\bibnamefont {Vaikuntanathan}},\ }\bibfield
  {title} {\bibinfo {title} {How dissipation constrains fluctuations in
  nonequilibrium liquids: Diffusion, structure, and biased interactions},\
  }\href {https://doi.org/10.1103/PhysRevX.9.041026} {\bibfield  {journal}
  {\bibinfo  {journal} {Phys. Rev. X}\ }\textbf {\bibinfo {volume} {9}},\
  \bibinfo {pages} {041026} (\bibinfo {year} {2019})}\BibitemShut {NoStop}%
\bibitem [{\citenamefont {Das}\ and\ \citenamefont
  {Limmer}(2023)}]{Limmer2023}%
  \BibitemOpen
  \bibfield  {author} {\bibinfo {author} {\bibfnamefont {A.}~\bibnamefont
  {Das}}\ and\ \bibinfo {author} {\bibfnamefont {D.~T.}\ \bibnamefont
  {Limmer}},\ }\bibfield  {title} {\bibinfo {title} {Nonequilibrium design
  strategies for functional colloidal assemblies},\ }\href
  {https://doi.org/10.1073/pnas.2217242120} {\bibfield  {journal} {\bibinfo
  {journal} {Proc. Natl. Acad. Sci. USA}\ }\textbf {\bibinfo {volume} {120}},\
  \bibinfo {pages} {e2217242120} (\bibinfo {year} {2023})}\BibitemShut
  {NoStop}%
\bibitem [{\citenamefont {Lamtyugina}\ \emph {et~al.}(2022)\citenamefont
  {Lamtyugina}, \citenamefont {Qiu}, \citenamefont {Fodor}, \citenamefont
  {Dinner},\ and\ \citenamefont {Vaikuntanathan}}]{CytoskeletonLD}%
  \BibitemOpen
  \bibfield  {author} {\bibinfo {author} {\bibfnamefont {A.}~\bibnamefont
  {Lamtyugina}}, \bibinfo {author} {\bibfnamefont {Y.}~\bibnamefont {Qiu}},
  \bibinfo {author} {\bibfnamefont {E.}~\bibnamefont {Fodor}}, \bibinfo
  {author} {\bibfnamefont {A.~R.}\ \bibnamefont {Dinner}},\ and\ \bibinfo
  {author} {\bibfnamefont {S.}~\bibnamefont {Vaikuntanathan}},\ }\bibfield
  {title} {\bibinfo {title} {Thermodynamic control of activity patterns in
  cytoskeletal networks},\ }\href
  {https://doi.org/10.1103/PhysRevLett.129.128002} {\bibfield  {journal}
  {\bibinfo  {journal} {Phys. Rev. Lett.}\ }\textbf {\bibinfo {volume} {129}},\
  \bibinfo {pages} {128002} (\bibinfo {year} {2022})}\BibitemShut {NoStop}%
\bibitem [{\citenamefont {Fodor}\ \emph {et~al.}(2022)\citenamefont {Fodor},
  \citenamefont {Jack},\ and\ \citenamefont {Cates}}]{Jack2022}%
  \BibitemOpen
  \bibfield  {author} {\bibinfo {author} {\bibfnamefont {E.}~\bibnamefont
  {Fodor}}, \bibinfo {author} {\bibfnamefont {R.~L.}\ \bibnamefont {Jack}},\
  and\ \bibinfo {author} {\bibfnamefont {M.~E.}\ \bibnamefont {Cates}},\
  }\bibfield  {title} {\bibinfo {title} {Irreversibility and biased ensembles
  in active matter: Insights from stochastic thermodynamics},\ }\href
  {https://doi.org/10.1146/annurev-conmatphys-031720-032419} {\bibfield
  {journal} {\bibinfo  {journal} {Annu. Rev. Condens. Matter Phys.}\ }\textbf
  {\bibinfo {volume} {13}},\ \bibinfo {pages} {215} (\bibinfo {year}
  {2022})}\BibitemShut {NoStop}%
\bibitem [{\citenamefont {Cagnetta}\ \emph {et~al.}(2017)\citenamefont
  {Cagnetta}, \citenamefont {Corberi}, \citenamefont {Gonnella},\ and\
  \citenamefont {Suma}}]{ABPLD0}%
  \BibitemOpen
  \bibfield  {author} {\bibinfo {author} {\bibfnamefont {F.}~\bibnamefont
  {Cagnetta}}, \bibinfo {author} {\bibfnamefont {F.}~\bibnamefont {Corberi}},
  \bibinfo {author} {\bibfnamefont {G.}~\bibnamefont {Gonnella}},\ and\
  \bibinfo {author} {\bibfnamefont {A.}~\bibnamefont {Suma}},\ }\bibfield
  {title} {\bibinfo {title} {Large fluctuations and dynamic phase transition in
  a system of self-propelled particles},\ }\href
  {https://doi.org/10.1103/PhysRevLett.119.158002} {\bibfield  {journal}
  {\bibinfo  {journal} {Phys. Rev. Lett.}\ }\textbf {\bibinfo {volume} {119}},\
  \bibinfo {pages} {158002} (\bibinfo {year} {2017})}\BibitemShut {NoStop}%
\bibitem [{\citenamefont {Whitelam}\ \emph {et~al.}(2018)\citenamefont
  {Whitelam}, \citenamefont {Klymko},\ and\ \citenamefont
  {Mandal}}]{Whitelam2018}%
  \BibitemOpen
  \bibfield  {author} {\bibinfo {author} {\bibfnamefont {S.}~\bibnamefont
  {Whitelam}}, \bibinfo {author} {\bibfnamefont {K.}~\bibnamefont {Klymko}},\
  and\ \bibinfo {author} {\bibfnamefont {D.}~\bibnamefont {Mandal}},\
  }\bibfield  {title} {\bibinfo {title} {{Phase separation and large deviations
  of lattice active matter}},\ }\href {https://doi.org/10.1063/1.5023403}
  {\bibfield  {journal} {\bibinfo  {journal} {J. Chem. Phys.}\ }\textbf
  {\bibinfo {volume} {148}},\ \bibinfo {pages} {154902} (\bibinfo {year}
  {2018})}\BibitemShut {NoStop}%
\bibitem [{\citenamefont {GrandPre}\ \emph {et~al.}(2021)\citenamefont
  {GrandPre}, \citenamefont {Klymko}, \citenamefont {Mandadapu},\ and\
  \citenamefont {Limmer}}]{GrandPre2021}%
  \BibitemOpen
  \bibfield  {author} {\bibinfo {author} {\bibfnamefont {T.}~\bibnamefont
  {GrandPre}}, \bibinfo {author} {\bibfnamefont {K.}~\bibnamefont {Klymko}},
  \bibinfo {author} {\bibfnamefont {K.~K.}\ \bibnamefont {Mandadapu}},\ and\
  \bibinfo {author} {\bibfnamefont {D.~T.}\ \bibnamefont {Limmer}},\ }\bibfield
   {title} {\bibinfo {title} {Entropy production fluctuations encode collective
  behavior in active matter},\ }\href
  {https://doi.org/10.1103/PhysRevE.103.012613} {\bibfield  {journal} {\bibinfo
   {journal} {Phys. Rev. E}\ }\textbf {\bibinfo {volume} {103}},\ \bibinfo
  {pages} {012613} (\bibinfo {year} {2021})}\BibitemShut {NoStop}%
\bibitem [{\citenamefont {Nemoto}\ \emph {et~al.}(2019)\citenamefont {Nemoto},
  \citenamefont {Fodor}, \citenamefont {Cates}, \citenamefont {Jack},\ and\
  \citenamefont {Tailleur}}]{ABPLD1}%
  \BibitemOpen
  \bibfield  {author} {\bibinfo {author} {\bibfnamefont {T.}~\bibnamefont
  {Nemoto}}, \bibinfo {author} {\bibfnamefont {E.}~\bibnamefont {Fodor}},
  \bibinfo {author} {\bibfnamefont {M.~E.}\ \bibnamefont {Cates}}, \bibinfo
  {author} {\bibfnamefont {R.~L.}\ \bibnamefont {Jack}},\ and\ \bibinfo
  {author} {\bibfnamefont {J.}~\bibnamefont {Tailleur}},\ }\bibfield  {title}
  {\bibinfo {title} {Optimizing active work: Dynamical phase transitions,
  collective motion, and jamming},\ }\href
  {https://doi.org/10.1103/PhysRevE.99.022605} {\bibfield  {journal} {\bibinfo
  {journal} {Phys. Rev. E}\ }\textbf {\bibinfo {volume} {99}},\ \bibinfo
  {pages} {022605} (\bibinfo {year} {2019})}\BibitemShut {NoStop}%
\bibitem [{\citenamefont {Keta}\ \emph {et~al.}(2021)\citenamefont {Keta},
  \citenamefont {Fodor}, \citenamefont {van Wijland}, \citenamefont {Cates},\
  and\ \citenamefont {Jack}}]{ABPLD2}%
  \BibitemOpen
  \bibfield  {author} {\bibinfo {author} {\bibfnamefont {Y.-E.}\ \bibnamefont
  {Keta}}, \bibinfo {author} {\bibfnamefont {E.}~\bibnamefont {Fodor}},
  \bibinfo {author} {\bibfnamefont {F.}~\bibnamefont {van Wijland}}, \bibinfo
  {author} {\bibfnamefont {M.~E.}\ \bibnamefont {Cates}},\ and\ \bibinfo
  {author} {\bibfnamefont {R.~L.}\ \bibnamefont {Jack}},\ }\bibfield  {title}
  {\bibinfo {title} {Collective motion in large deviations of active
  particles},\ }\href {https://doi.org/10.1103/PhysRevE.103.022603} {\bibfield
  {journal} {\bibinfo  {journal} {Phys. Rev. E}\ }\textbf {\bibinfo {volume}
  {103}},\ \bibinfo {pages} {022603} (\bibinfo {year} {2021})}\BibitemShut
  {NoStop}%
\bibitem [{\citenamefont {Agranov}\ \emph {et~al.}(2023)\citenamefont
  {Agranov}, \citenamefont {Cates},\ and\ \citenamefont {Jack}}]{Agranov2023}%
  \BibitemOpen
  \bibfield  {author} {\bibinfo {author} {\bibfnamefont {T.}~\bibnamefont
  {Agranov}}, \bibinfo {author} {\bibfnamefont {M.~E.}\ \bibnamefont {Cates}},\
  and\ \bibinfo {author} {\bibfnamefont {R.~L.}\ \bibnamefont {Jack}},\
  }\bibfield  {title} {\bibinfo {title} {Tricritical behavior in dynamical
  phase transitions},\ }\href {https://doi.org/10.1103/PhysRevLett.131.017102}
  {\bibfield  {journal} {\bibinfo  {journal} {Phys. Rev. Lett.}\ }\textbf
  {\bibinfo {volume} {131}},\ \bibinfo {pages} {017102} (\bibinfo {year}
  {2023})}\BibitemShut {NoStop}%
\bibitem [{\citenamefont {Caprini}\ \emph {et~al.}(2020)\citenamefont
  {Caprini}, \citenamefont {Marini Bettolo~Marconi},\ and\ \citenamefont
  {Puglisi}}]{alignment_nosync1}%
  \BibitemOpen
  \bibfield  {author} {\bibinfo {author} {\bibfnamefont {L.}~\bibnamefont
  {Caprini}}, \bibinfo {author} {\bibfnamefont {U.}~\bibnamefont {Marini
  Bettolo~Marconi}},\ and\ \bibinfo {author} {\bibfnamefont {A.}~\bibnamefont
  {Puglisi}},\ }\bibfield  {title} {\bibinfo {title} {Spontaneous velocity
  alignment in motility-induced phase separation},\ }\href
  {https://doi.org/10.1103/PhysRevLett.124.078001} {\bibfield  {journal}
  {\bibinfo  {journal} {Phys. Rev. Lett.}\ }\textbf {\bibinfo {volume} {124}},\
  \bibinfo {pages} {078001} (\bibinfo {year} {2020})}\BibitemShut {NoStop}%
\bibitem [{\citenamefont {Caprini}\ \emph {et~al.}(2021)\citenamefont
  {Caprini}, \citenamefont {Maggi},\ and\ \citenamefont {Marini
  Bettolo~Marconi}}]{alignment_ring_nosync2}%
  \BibitemOpen
  \bibfield  {author} {\bibinfo {author} {\bibfnamefont {L.}~\bibnamefont
  {Caprini}}, \bibinfo {author} {\bibfnamefont {C.}~\bibnamefont {Maggi}},\
  and\ \bibinfo {author} {\bibfnamefont {U.}~\bibnamefont {Marini
  Bettolo~Marconi}},\ }\bibfield  {title} {\bibinfo {title} {{Collective
  effects in confined active Brownian particles}},\ }\href
  {https://doi.org/10.1063/5.0051315} {\bibfield  {journal} {\bibinfo
  {journal} {The Journal of Chemical Physics}\ }\textbf {\bibinfo {volume}
  {154}},\ \bibinfo {pages} {244901} (\bibinfo {year} {2021})}\BibitemShut
  {NoStop}%
\bibitem [{\citenamefont {Casiulis}\ and\ \citenamefont
  {Levine}(2022)}]{flocking_noalignment}%
  \BibitemOpen
  \bibfield  {author} {\bibinfo {author} {\bibfnamefont {M.}~\bibnamefont
  {Casiulis}}\ and\ \bibinfo {author} {\bibfnamefont {D.}~\bibnamefont
  {Levine}},\ }\bibfield  {title} {\bibinfo {title} {Emergent synchronization
  and flocking in purely repulsive self-navigating particles},\ }\href
  {https://doi.org/10.1103/PhysRevE.106.044611} {\bibfield  {journal} {\bibinfo
   {journal} {Phys. Rev. E}\ }\textbf {\bibinfo {volume} {106}},\ \bibinfo
  {pages} {044611} (\bibinfo {year} {2022})}\BibitemShut {NoStop}%
\bibitem [{SM()}]{SM}%
  \BibitemOpen
  \href@noop {} {}\bibinfo {note} {See supplemental material at
  \url{URL_by_publisher}, which includes Refs.~\cite{Lecomte_2007,
  weighed_voronoi, freud_package} for movies and details on numerical
  simulations.}\BibitemShut {Stop}%
\bibitem [{\citenamefont {Wittich}\ and\ \citenamefont
  {Deiters}(2010)}]{CrystalAssemblyBoxEffects}%
  \BibitemOpen
  \bibfield  {author} {\bibinfo {author} {\bibfnamefont {B.}~\bibnamefont
  {Wittich}}\ and\ \bibinfo {author} {\bibfnamefont {U.~K.}\ \bibnamefont
  {Deiters}},\ }\bibfield  {title} {\bibinfo {title} {The influence of the
  simulation box geometry in solid-state molecular simulations: phase behaviour
  of lithium iodide in a dynamic monte carlo simulation},\ }\href
  {https://doi.org/10.1080/08927020903483320} {\bibfield  {journal} {\bibinfo
  {journal} {Mol. Simul.}\ }\textbf {\bibinfo {volume} {36}},\ \bibinfo {pages}
  {364} (\bibinfo {year} {2010})}\BibitemShut {NoStop}%
\bibitem [{\citenamefont {Zhang}\ \emph {et~al.}(2013)\citenamefont {Zhang},
  \citenamefont {Stillinger},\ and\ \citenamefont
  {Torquato}}]{TorquatoCrystalAssembly}%
  \BibitemOpen
  \bibfield  {author} {\bibinfo {author} {\bibfnamefont {G.}~\bibnamefont
  {Zhang}}, \bibinfo {author} {\bibfnamefont {F.~H.}\ \bibnamefont
  {Stillinger}},\ and\ \bibinfo {author} {\bibfnamefont {S.}~\bibnamefont
  {Torquato}},\ }\bibfield  {title} {\bibinfo {title} {Probing the limitations
  of isotropic pair potentials to produce ground-state structural extremes via
  inverse statistical mechanics},\ }\href
  {https://doi.org/10.1103/PhysRevE.88.042309} {\bibfield  {journal} {\bibinfo
  {journal} {Phys. Rev. E}\ }\textbf {\bibinfo {volume} {88}},\ \bibinfo
  {pages} {042309} (\bibinfo {year} {2013})}\BibitemShut {NoStop}%
\bibitem [{\citenamefont {Piñeros}\ and\ \citenamefont
  {Truskett}(2017)}]{OpenSelfAssembly}%
  \BibitemOpen
  \bibfield  {author} {\bibinfo {author} {\bibfnamefont {W.~D.}\ \bibnamefont
  {Piñeros}}\ and\ \bibinfo {author} {\bibfnamefont {T.~M.}\ \bibnamefont
  {Truskett}},\ }\bibfield  {title} {\bibinfo {title} {{Designing pairwise
  interactions that stabilize open crystals: Trunc. square and trunc. hex.
  lattices}},\ }\href {https://doi.org/10.1063/1.4979715} {\bibfield  {journal}
  {\bibinfo  {journal} {J. Chem. Phys.}\ }\textbf {\bibinfo {volume} {146}},\
  \bibinfo {pages} {144501} (\bibinfo {year} {2017})}\BibitemShut {NoStop}%
\bibitem [{\citenamefont {Giardin\`a}\ \emph {et~al.}(2006)\citenamefont
  {Giardin\`a}, \citenamefont {Kurchan},\ and\ \citenamefont
  {Peliti}}]{CloningMethod}%
  \BibitemOpen
  \bibfield  {author} {\bibinfo {author} {\bibfnamefont {C.}~\bibnamefont
  {Giardin\`a}}, \bibinfo {author} {\bibfnamefont {J.}~\bibnamefont
  {Kurchan}},\ and\ \bibinfo {author} {\bibfnamefont {L.}~\bibnamefont
  {Peliti}},\ }\bibfield  {title} {\bibinfo {title} {Direct eval. of
  large-deviation functions},\ }\href
  {https://doi.org/10.1103/PhysRevLett.96.120603} {\bibfield  {journal}
  {\bibinfo  {journal} {Phys. Rev. Lett.}\ }\textbf {\bibinfo {volume} {96}},\
  \bibinfo {pages} {120603} (\bibinfo {year} {2006})}\BibitemShut {NoStop}%
\bibitem [{\citenamefont {Nemoto}\ \emph {et~al.}(2016)\citenamefont {Nemoto},
  \citenamefont {Bouchet}, \citenamefont {Jack},\ and\ \citenamefont
  {Lecomte}}]{Nemoto2016}%
  \BibitemOpen
  \bibfield  {author} {\bibinfo {author} {\bibfnamefont {T.}~\bibnamefont
  {Nemoto}}, \bibinfo {author} {\bibfnamefont {F.}~\bibnamefont {Bouchet}},
  \bibinfo {author} {\bibfnamefont {R.~L.}\ \bibnamefont {Jack}},\ and\
  \bibinfo {author} {\bibfnamefont {V.}~\bibnamefont {Lecomte}},\ }\bibfield
  {title} {\bibinfo {title} {Population-dynamics method with a multicanonical
  feedback control},\ }\href {https://doi.org/10.1103/PhysRevE.93.062123}
  {\bibfield  {journal} {\bibinfo  {journal} {Phys. Rev. E}\ }\textbf {\bibinfo
  {volume} {93}},\ \bibinfo {pages} {062123} (\bibinfo {year}
  {2016})}\BibitemShut {NoStop}%
\bibitem [{\citenamefont {Hurtado-Guti\'errez}\ \emph
  {et~al.}(2020)\citenamefont {Hurtado-Guti\'errez}, \citenamefont {Carollo},
  \citenamefont {P\'erez-Espigares},\ and\ \citenamefont
  {Hurtado}}]{TimeCrystalsKuramotoLD}%
  \BibitemOpen
  \bibfield  {author} {\bibinfo {author} {\bibfnamefont {R.}~\bibnamefont
  {Hurtado-Guti\'errez}}, \bibinfo {author} {\bibfnamefont {F.}~\bibnamefont
  {Carollo}}, \bibinfo {author} {\bibfnamefont {C.}~\bibnamefont
  {P\'erez-Espigares}},\ and\ \bibinfo {author} {\bibfnamefont {P.~I.}\
  \bibnamefont {Hurtado}},\ }\bibfield  {title} {\bibinfo {title} {Building
  continuous time crystals from rare events},\ }\href
  {https://doi.org/10.1103/PhysRevLett.125.160601} {\bibfield  {journal}
  {\bibinfo  {journal} {Phys. Rev. Lett.}\ }\textbf {\bibinfo {volume} {125}},\
  \bibinfo {pages} {160601} (\bibinfo {year} {2020})}\BibitemShut {NoStop}%
\bibitem [{\citenamefont {Manacorda}\ and\ \citenamefont
  {Fodor}(2023)}]{Manacorda2023}%
  \BibitemOpen
  \bibfield  {author} {\bibinfo {author} {\bibfnamefont {A.}~\bibnamefont
  {Manacorda}}\ and\ \bibinfo {author} {\bibfnamefont {E.}~\bibnamefont
  {Fodor}},\ }\bibfield  {title} {\bibinfo {title} {Pulsating with discrete
  symmetry},\ }\href@noop {} {\bibfield  {journal} {\bibinfo  {journal} {ArXiv
  e-prints}\ } (\bibinfo {year} {2023})},\ \Eprint
  {https://arxiv.org/abs/2310.14370} {arXiv:2310.14370} \BibitemShut {NoStop}%
\bibitem [{\citenamefont {Briand}\ \emph {et~al.}(2018)\citenamefont {Briand},
  \citenamefont {Schindler},\ and\ \citenamefont
  {Dauchot}}]{SPPHexFluidization}%
  \BibitemOpen
  \bibfield  {author} {\bibinfo {author} {\bibfnamefont {G.}~\bibnamefont
  {Briand}}, \bibinfo {author} {\bibfnamefont {M.}~\bibnamefont {Schindler}},\
  and\ \bibinfo {author} {\bibfnamefont {O.}~\bibnamefont {Dauchot}},\
  }\bibfield  {title} {\bibinfo {title} {Spontaneously flowing crystal of
  self-propelled particles},\ }\href
  {https://doi.org/10.1103/PhysRevLett.120.208001} {\bibfield  {journal}
  {\bibinfo  {journal} {Phys. Rev. Lett.}\ }\textbf {\bibinfo {volume} {120}},\
  \bibinfo {pages} {208001} (\bibinfo {year} {2018})}\BibitemShut {NoStop}%
\bibitem [{\citenamefont {Berthier}\ \emph {et~al.}(2019)\citenamefont
  {Berthier}, \citenamefont {Flenner},\ and\ \citenamefont
  {Szamel}}]{glassyAM}%
  \BibitemOpen
  \bibfield  {author} {\bibinfo {author} {\bibfnamefont {L.}~\bibnamefont
  {Berthier}}, \bibinfo {author} {\bibfnamefont {E.}~\bibnamefont {Flenner}},\
  and\ \bibinfo {author} {\bibfnamefont {G.}~\bibnamefont {Szamel}},\
  }\bibfield  {title} {\bibinfo {title} {{Glassy dynamics in dense systems of
  active particles}},\ }\href {https://doi.org/10.1063/1.5093240} {\bibfield
  {journal} {\bibinfo  {journal} {J. Chem. Phys.}\ }\textbf {\bibinfo {volume}
  {150}},\ \bibinfo {pages} {200901} (\bibinfo {year} {2019})}\BibitemShut
  {NoStop}%
\bibitem [{\citenamefont {Mandal}\ \emph {et~al.}(2020)\citenamefont {Mandal},
  \citenamefont {Bhuyan}, \citenamefont {Chaudhuri}, \citenamefont {Dasgupta},\
  and\ \citenamefont {Rao}}]{PlasticAM}%
  \BibitemOpen
  \bibfield  {author} {\bibinfo {author} {\bibfnamefont {R.}~\bibnamefont
  {Mandal}}, \bibinfo {author} {\bibfnamefont {P.~J.}\ \bibnamefont {Bhuyan}},
  \bibinfo {author} {\bibfnamefont {P.}~\bibnamefont {Chaudhuri}}, \bibinfo
  {author} {\bibfnamefont {C.}~\bibnamefont {Dasgupta}},\ and\ \bibinfo
  {author} {\bibfnamefont {M.}~\bibnamefont {Rao}},\ }\bibfield  {title}
  {\bibinfo {title} {Extreme active matter at high densities},\ }\href
  {https://doi.org/10.1038/s41467-020-16130-x} {\bibfield  {journal} {\bibinfo
  {journal} {Nat. Commun.}\ }\textbf {\bibinfo {volume} {11}},\ \bibinfo
  {pages} {2581} (\bibinfo {year} {2020})}\BibitemShut {NoStop}%
\bibitem [{\citenamefont {Morse}\ \emph {et~al.}(2021)\citenamefont {Morse},
  \citenamefont {Roy}, \citenamefont {Agoritsas}, \citenamefont {Stanifer},
  \citenamefont {Corwin},\ and\ \citenamefont {Manning}}]{Manning2021}%
  \BibitemOpen
  \bibfield  {author} {\bibinfo {author} {\bibfnamefont {P.~K.}\ \bibnamefont
  {Morse}}, \bibinfo {author} {\bibfnamefont {S.}~\bibnamefont {Roy}}, \bibinfo
  {author} {\bibfnamefont {E.}~\bibnamefont {Agoritsas}}, \bibinfo {author}
  {\bibfnamefont {E.}~\bibnamefont {Stanifer}}, \bibinfo {author}
  {\bibfnamefont {E.~I.}\ \bibnamefont {Corwin}},\ and\ \bibinfo {author}
  {\bibfnamefont {M.~L.}\ \bibnamefont {Manning}},\ }\bibfield  {title}
  {\bibinfo {title} {A direct link between active matter and sheared granular
  systems},\ }\href {https://doi.org/10.1073/pnas.2019909118} {\bibfield
  {journal} {\bibinfo  {journal} {Proc. Natl. Acad. Sci. USA}\ }\textbf
  {\bibinfo {volume} {118}},\ \bibinfo {pages} {e2019909118} (\bibinfo {year}
  {2021})}\BibitemShut {NoStop}%
\bibitem [{\citenamefont {Tjhung}\ and\ \citenamefont
  {Berthier}(2020)}]{GrowingAMGlassAnalogies}%
  \BibitemOpen
  \bibfield  {author} {\bibinfo {author} {\bibfnamefont {E.}~\bibnamefont
  {Tjhung}}\ and\ \bibinfo {author} {\bibfnamefont {L.}~\bibnamefont
  {Berthier}},\ }\bibfield  {title} {\bibinfo {title} {Analogies between
  growing dense active matter and soft driven glasses},\ }\href
  {https://doi.org/10.1103/PhysRevResearch.2.043334} {\bibfield  {journal}
  {\bibinfo  {journal} {Phys. Rev. Res.}\ }\textbf {\bibinfo {volume} {2}},\
  \bibinfo {pages} {043334} (\bibinfo {year} {2020})}\BibitemShut {NoStop}%
\bibitem [{\citenamefont {Ninarello}\ \emph {et~al.}(2017)\citenamefont
  {Ninarello}, \citenamefont {Berthier},\ and\ \citenamefont
  {Coslovich}}]{Berthier2017}%
  \BibitemOpen
  \bibfield  {author} {\bibinfo {author} {\bibfnamefont {A.}~\bibnamefont
  {Ninarello}}, \bibinfo {author} {\bibfnamefont {L.}~\bibnamefont
  {Berthier}},\ and\ \bibinfo {author} {\bibfnamefont {D.}~\bibnamefont
  {Coslovich}},\ }\bibfield  {title} {\bibinfo {title} {Models and algorithms
  for the next generation of glass transition studies},\ }\href
  {https://doi.org/10.1103/PhysRevX.7.021039} {\bibfield  {journal} {\bibinfo
  {journal} {Phys. Rev. X}\ }\textbf {\bibinfo {volume} {7}},\ \bibinfo {pages}
  {021039} (\bibinfo {year} {2017})}\BibitemShut {NoStop}%
\bibitem [{\citenamefont {Brito}\ \emph {et~al.}(2018)\citenamefont {Brito},
  \citenamefont {Lerner},\ and\ \citenamefont {Wyart}}]{SWAPglass}%
  \BibitemOpen
  \bibfield  {author} {\bibinfo {author} {\bibfnamefont {C.}~\bibnamefont
  {Brito}}, \bibinfo {author} {\bibfnamefont {E.}~\bibnamefont {Lerner}},\ and\
  \bibinfo {author} {\bibfnamefont {M.}~\bibnamefont {Wyart}},\ }\bibfield
  {title} {\bibinfo {title} {Theory for swap acceleration near the glass and
  jamming transitions for continuously polydisperse particles},\ }\href
  {https://doi.org/10.1103/PhysRevX.8.031050} {\bibfield  {journal} {\bibinfo
  {journal} {Phys. Rev. X}\ }\textbf {\bibinfo {volume} {8}},\ \bibinfo {pages}
  {031050} (\bibinfo {year} {2018})}\BibitemShut {NoStop}%
\bibitem [{\citenamefont {Hagh}\ \emph {et~al.}(2022)\citenamefont {Hagh},
  \citenamefont {Nagel}, \citenamefont {Liu}, \citenamefont {Manning},\ and\
  \citenamefont {Corwin}}]{particle_size_change_jamming}%
  \BibitemOpen
  \bibfield  {author} {\bibinfo {author} {\bibfnamefont {V.~F.}\ \bibnamefont
  {Hagh}}, \bibinfo {author} {\bibfnamefont {S.~R.}\ \bibnamefont {Nagel}},
  \bibinfo {author} {\bibfnamefont {A.~J.}\ \bibnamefont {Liu}}, \bibinfo
  {author} {\bibfnamefont {M.~L.}\ \bibnamefont {Manning}},\ and\ \bibinfo
  {author} {\bibfnamefont {E.~I.}\ \bibnamefont {Corwin}},\ }\bibfield  {title}
  {\bibinfo {title} {Transient learning degrees of freedom for introducing
  function in materials},\ }\href {https://doi.org/10.1073/pnas.2117622119}
  {\bibfield  {journal} {\bibinfo  {journal} {Proc. Natl. Acad. of Sci. USA}\
  }\textbf {\bibinfo {volume} {119}},\ \bibinfo {pages} {e2117622119} (\bibinfo
  {year} {2022})}\BibitemShut {NoStop}%
\bibitem [{\citenamefont {Bolton-Lum}\ \emph {et~al.}(2024)\citenamefont
  {Bolton-Lum}, \citenamefont {Dennis}, \citenamefont {Morse},\ and\
  \citenamefont {Corwin}}]{ideal_glass_radii_change_method}%
  \BibitemOpen
  \bibfield  {author} {\bibinfo {author} {\bibfnamefont {V.}~\bibnamefont
  {Bolton-Lum}}, \bibinfo {author} {\bibfnamefont {R.~C.}\ \bibnamefont
  {Dennis}}, \bibinfo {author} {\bibfnamefont {P.}~\bibnamefont {Morse}},\ and\
  \bibinfo {author} {\bibfnamefont {E.}~\bibnamefont {Corwin}},\ }\href
  {https://arxiv.org/abs/2404.07492} {\bibinfo {title} {The ideal glass and the
  ideal disk packing in two dimensions}} (\bibinfo {year} {2024}),\ \Eprint
  {https://arxiv.org/abs/2404.07492} {arXiv:2404.07492 [cond-mat.soft]}
  \BibitemShut {NoStop}%
\bibitem [{\citenamefont {Manning}(2023)}]{Manning2023}%
  \BibitemOpen
  \bibfield  {author} {\bibinfo {author} {\bibfnamefont {M.~L.}\ \bibnamefont
  {Manning}},\ }\bibfield  {title} {\bibinfo {title} {Essay: Collections of
  deformable particles present exciting challenges for soft matter and
  biological physics},\ }\href {https://doi.org/10.1103/PhysRevLett.130.130002}
  {\bibfield  {journal} {\bibinfo  {journal} {Phys. Rev. Lett.}\ }\textbf
  {\bibinfo {volume} {130}},\ \bibinfo {pages} {130002} (\bibinfo {year}
  {2023})}\BibitemShut {NoStop}%
\bibitem [{\citenamefont {Sakaguchi}\ and\ \citenamefont
  {Kuramoto}(1986)}]{Kuramoto1986}%
  \BibitemOpen
  \bibfield  {author} {\bibinfo {author} {\bibfnamefont {H.}~\bibnamefont
  {Sakaguchi}}\ and\ \bibinfo {author} {\bibfnamefont {Y.}~\bibnamefont
  {Kuramoto}},\ }\bibfield  {title} {\bibinfo {title} {{A Soluble Active
  Rotater Model Showing Phase Transitions via Mutual Entertainment}},\ }\href
  {https://doi.org/10.1143/PTP.76.576} {\bibfield  {journal} {\bibinfo
  {journal} {Prog. Theor. Phys.}\ }\textbf {\bibinfo {volume} {76}},\ \bibinfo
  {pages} {576} (\bibinfo {year} {1986})}\BibitemShut {NoStop}%
\bibitem [{\citenamefont {O'Keeffe}\ \emph {et~al.}(2017)\citenamefont
  {O'Keeffe}, \citenamefont {Hong},\ and\ \citenamefont
  {Strogatz}}]{Strogatz2017}%
  \BibitemOpen
  \bibfield  {author} {\bibinfo {author} {\bibfnamefont {K.~P.}\ \bibnamefont
  {O'Keeffe}}, \bibinfo {author} {\bibfnamefont {H.}~\bibnamefont {Hong}},\
  and\ \bibinfo {author} {\bibfnamefont {S.~H.}\ \bibnamefont {Strogatz}},\
  }\bibfield  {title} {\bibinfo {title} {Oscillators that sync and swarm},\
  }\href {https://doi.org/10.1038/s41467-017-01190-3} {\bibfield  {journal}
  {\bibinfo  {journal} {Nat. Commun.}\ }\textbf {\bibinfo {volume} {8}},\
  \bibinfo {pages} {1504} (\bibinfo {year} {2017})}\BibitemShut {NoStop}%
\bibitem [{\citenamefont {Adorjani}\ \emph {et~al.}(2023)\citenamefont
  {Adorjani}, \citenamefont {Libal}, \citenamefont {Reichhardt},\ and\
  \citenamefont {Reichhardt}}]{Reichhardt2023}%
  \BibitemOpen
  \bibfield  {author} {\bibinfo {author} {\bibfnamefont {B.}~\bibnamefont
  {Adorjani}}, \bibinfo {author} {\bibfnamefont {A.}~\bibnamefont {Libal}},
  \bibinfo {author} {\bibfnamefont {C.}~\bibnamefont {Reichhardt}},\ and\
  \bibinfo {author} {\bibfnamefont {C.~J.~O.}\ \bibnamefont {Reichhardt}},\
  }\bibfield  {title} {\bibinfo {title} {Motility induced phase separation and
  frustration in active matter swarmalators},\ }\href@noop {} {\bibfield
  {journal} {\bibinfo  {journal} {ArXiv e-prints}\ } (\bibinfo {year}
  {2023})},\ \Eprint {https://arxiv.org/abs/2309.10937} {arXiv:2309.10937}
  \BibitemShut {NoStop}%
\bibitem [{\citenamefont {Whitelam}\ \emph {et~al.}(2020)\citenamefont
  {Whitelam}, \citenamefont {Jacobson},\ and\ \citenamefont
  {Tamblyn}}]{Whitelam2020}%
  \BibitemOpen
  \bibfield  {author} {\bibinfo {author} {\bibfnamefont {S.}~\bibnamefont
  {Whitelam}}, \bibinfo {author} {\bibfnamefont {D.}~\bibnamefont {Jacobson}},\
  and\ \bibinfo {author} {\bibfnamefont {I.}~\bibnamefont {Tamblyn}},\
  }\bibfield  {title} {\bibinfo {title} {{Evolutionary reinforcement learning
  of dynamical large deviations}},\ }\href {https://doi.org/10.1063/5.0015301}
  {\bibfield  {journal} {\bibinfo  {journal} {J. Chem. Phys.}\ }\textbf
  {\bibinfo {volume} {153}},\ \bibinfo {pages} {044113} (\bibinfo {year}
  {2020})}\BibitemShut {NoStop}%
\bibitem [{\citenamefont {Das}\ \emph {et~al.}(2021)\citenamefont {Das},
  \citenamefont {Rose}, \citenamefont {Garrahan},\ and\ \citenamefont
  {Limmer}}]{Garrahan2021}%
  \BibitemOpen
  \bibfield  {author} {\bibinfo {author} {\bibfnamefont {A.}~\bibnamefont
  {Das}}, \bibinfo {author} {\bibfnamefont {D.~C.}\ \bibnamefont {Rose}},
  \bibinfo {author} {\bibfnamefont {J.~P.}\ \bibnamefont {Garrahan}},\ and\
  \bibinfo {author} {\bibfnamefont {D.~T.}\ \bibnamefont {Limmer}},\ }\bibfield
   {title} {\bibinfo {title} {{Reinforcement learning of rare diffusive
  dynamics}},\ }\href {https://doi.org/10.1063/5.0057323} {\bibfield  {journal}
  {\bibinfo  {journal} {J. Chem. Phys.}\ }\textbf {\bibinfo {volume} {155}},\
  \bibinfo {pages} {134105} (\bibinfo {year} {2021})}\BibitemShut {NoStop}%
\bibitem [{\citenamefont {Yan}\ \emph {et~al.}(2022)\citenamefont {Yan},
  \citenamefont {Touchette},\ and\ \citenamefont {Rotskoff}}]{Rotskoff2022}%
  \BibitemOpen
  \bibfield  {author} {\bibinfo {author} {\bibfnamefont {J.}~\bibnamefont
  {Yan}}, \bibinfo {author} {\bibfnamefont {H.}~\bibnamefont {Touchette}},\
  and\ \bibinfo {author} {\bibfnamefont {G.~M.}\ \bibnamefont {Rotskoff}},\
  }\bibfield  {title} {\bibinfo {title} {Learning nonequilibrium control forces
  to characterize dynamical phase transitions},\ }\href
  {https://doi.org/10.1103/PhysRevE.105.024115} {\bibfield  {journal} {\bibinfo
   {journal} {Phys. Rev. E}\ }\textbf {\bibinfo {volume} {105}},\ \bibinfo
  {pages} {024115} (\bibinfo {year} {2022})}\BibitemShut {NoStop}%
\bibitem [{\citenamefont {Mickel}\ \emph {et~al.}(2013)\citenamefont {Mickel},
  \citenamefont {Kapfer}, \citenamefont {Schröder-Turk},\ and\ \citenamefont
  {Mecke}}]{weighed_voronoi}%
  \BibitemOpen
  \bibfield  {author} {\bibinfo {author} {\bibfnamefont {W.}~\bibnamefont
  {Mickel}}, \bibinfo {author} {\bibfnamefont {S.~C.}\ \bibnamefont {Kapfer}},
  \bibinfo {author} {\bibfnamefont {G.~E.}\ \bibnamefont {Schröder-Turk}},\
  and\ \bibinfo {author} {\bibfnamefont {K.}~\bibnamefont {Mecke}},\ }\bibfield
   {title} {\bibinfo {title} {{Shortcomings of the bond orientational order
  parameters for the analysis of disordered particulate matter}},\ }\href
  {https://doi.org/10.1063/1.4774084} {\bibfield  {journal} {\bibinfo
  {journal} {J. Chem. Phys.}\ }\textbf {\bibinfo {volume} {138}},\ \bibinfo
  {pages} {044501} (\bibinfo {year} {2013})}\BibitemShut {NoStop}%
\bibitem [{fre(2020)}]{freud_package}%
  \BibitemOpen
  \bibfield  {title} {\bibinfo {title} {freud: A software suite for high
  throughput analysis of particle simulation data},\ }\href
  {https://doi.org/https://doi.org/10.1016/j.cpc.2020.107275} {\bibfield
  {journal} {\bibinfo  {journal} {Comput. Phys. Commun.}\ }\textbf {\bibinfo
  {volume} {254}},\ \bibinfo {pages} {107275} (\bibinfo {year}
  {2020})}\BibitemShut {NoStop}%
\bibitem [{\citenamefont {Lecomte}\ and\ \citenamefont
  {Tailleur}(2007)}]{Lecomte_2007}%
  \BibitemOpen
  \bibfield  {author} {\bibinfo {author} {\bibfnamefont {V.}~\bibnamefont
  {Lecomte}}\ and\ \bibinfo {author} {\bibfnamefont {J.}~\bibnamefont
  {Tailleur}},\ }\bibfield  {title} {\bibinfo {title} {A numerical approach to
  large deviations in continuous time},\ }\href
  {https://doi.org/10.1088/1742-5468/2007/03/P03004} {\bibfield  {journal}
  {\bibinfo  {journal} {J. Stat. Mech.}\ }\textbf {\bibinfo {volume} {2007}},\
  \bibinfo {pages} {P03004} (\bibinfo {year} {2007})}\BibitemShut {NoStop}%
\end{thebibliography}
%

\end{document}


\title{Supplemental material: Biased ensembles of pulsating active matter}

\author{William D. Pi\~neros}
\author{\'Etienne Fodor}
\affiliation{Department of Physics and Materials Science, University of Luxembourg, L-1511 Luxembourg, Luxembourg}

\maketitle


\begin{figure*}
  \centering
  \includegraphics[scale=.18]{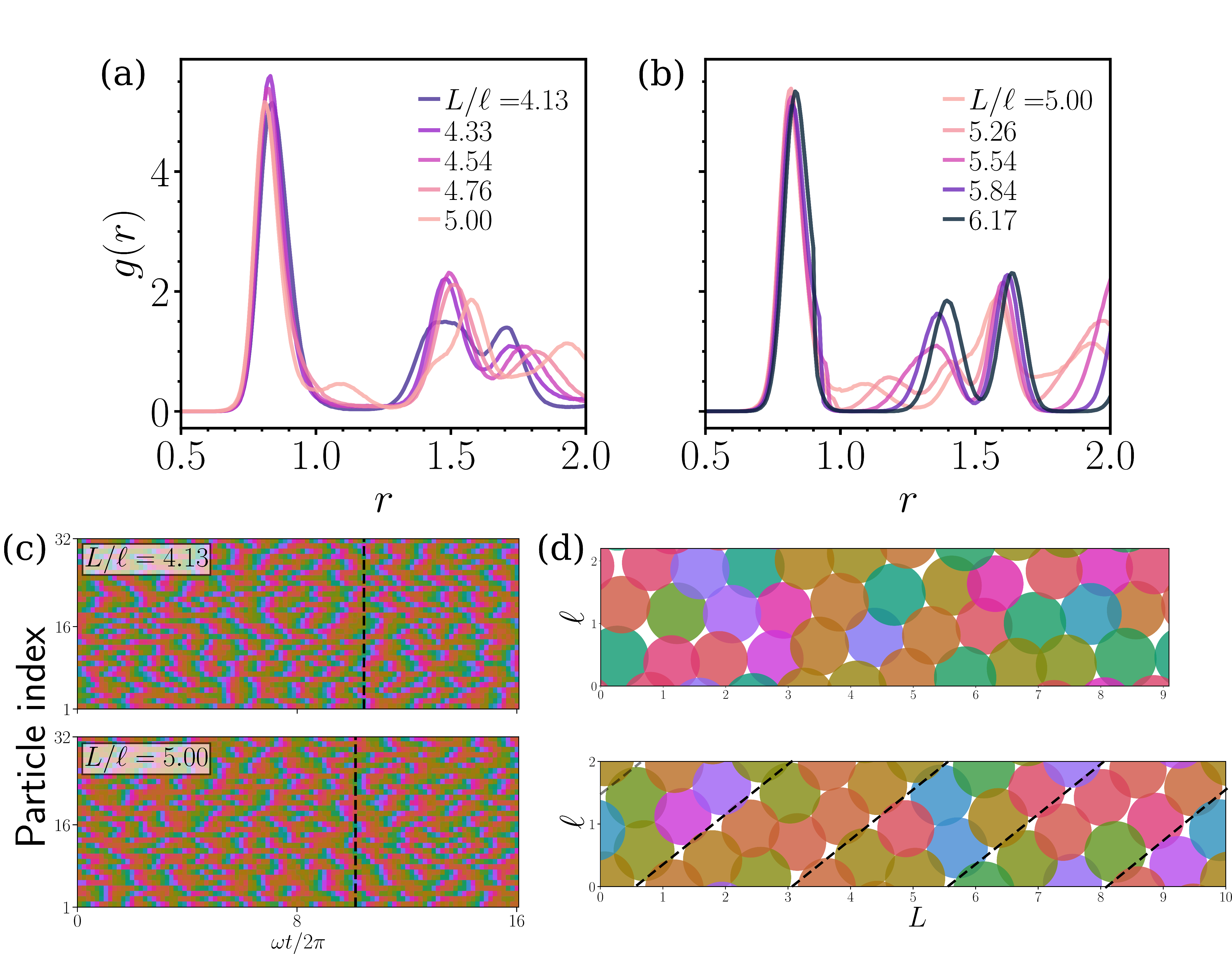}
  \caption{Radial distribution function for (a) $\Lol\in[4.13,5.00]$ and (b) $\Lol\in[5.00,6.17]$.
  (c)~Trajectory of unbiased dynamics and (d)~corresponding snapshots at the time denoted by the vertical dashed line: the case $L/\ell=5$ displays a regular packing configuration.
  Parameters: $N=32$, $\rho=1.6$.} 
\label{fig:s0_gr_all}
\end{figure*}

\section{Unbiased dynamics}
\subsection{Structural analysis}

To highlight the effect of box geometry in constraining the unbiased dynamics, we measure the radial distribution $g(r)$ [Eq.~(7) of main text] over all probed values of $\Lol$. Note that while equivalent radial sampling is limited to $r=\ell/2$, we nonetheless extend the sampling range up to $r=L/2$ by appropriately normalizing over an ideal gas distribution with same number of particles and box dimensions, which we sample numerically. This ensures we preserve correct relative weight behavior of the radial distribution as a function of distance. As shown in Figs.~\ref{fig:s0_gr_all}(a-b), systems with different $\Lol$ indeed display distinct distribution profiles. In particular, higher order coordination shells ($r>1$) manifest at different ranges and are indicative of the overall packing arrangement of each system.

We now illustrate this difference in packing structure for representative unbiased systems that result in cycles ($\Lol$=4.13), or arrest ($\Lol=5$) when biased. As seen in Figs.~\ref{fig:s0_gr_all}(c,d), a regular structure for $L/\ell=5$ already contrasts with the more irregular one at $\Lol=4.13$ (see also movies online). These results demonstrate how box geometry constrains the available configurations of the system and hence alter its average structure. In turn, such constraint affects the measurements of order and current [see $\phiAve$ and $\nuAve$ in Fig.~2 of main text] for different values of $\Lol$, and ultimately impacts the rare events of the dynamics.

\subsection{Scaling behavior of order parameter}

\begin{figure*}
  \centering
  \includegraphics[scale=.3]{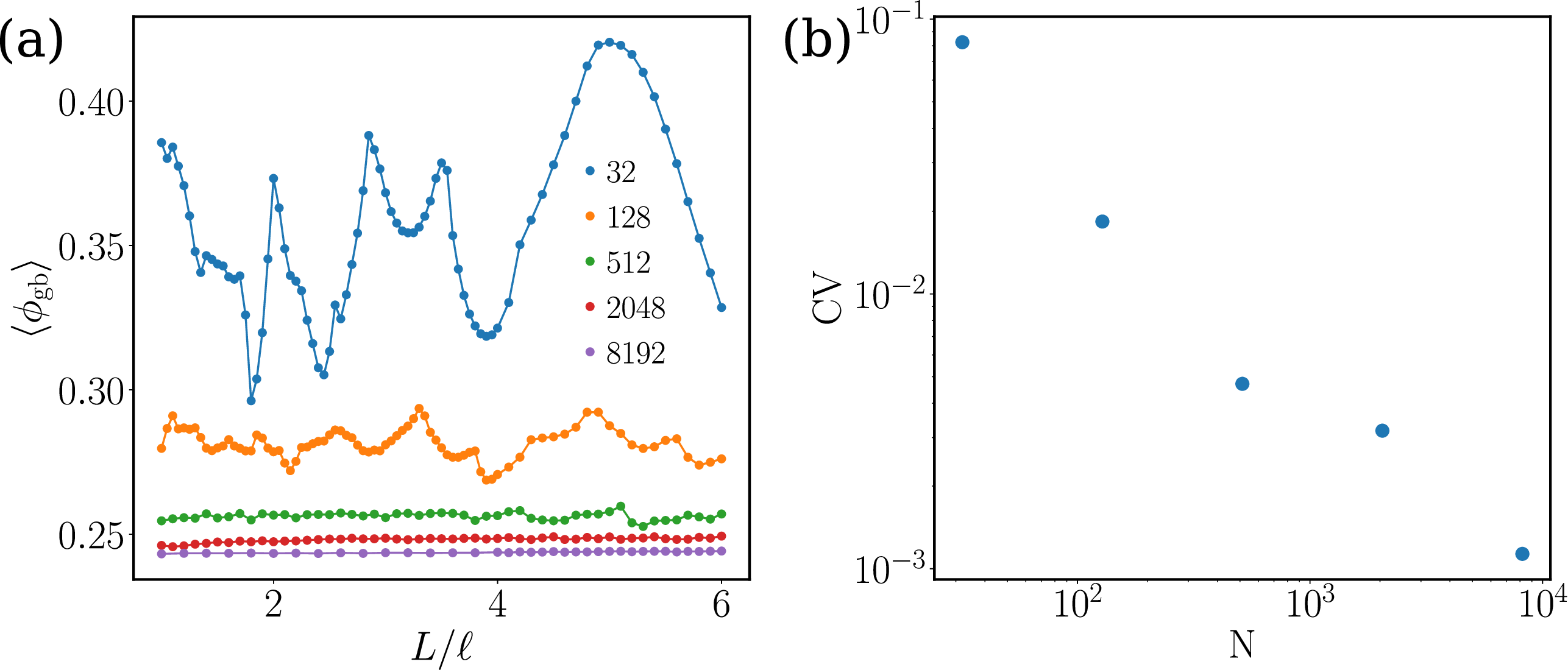}
  \caption{Scaling of the global order parameter $\phi_{\rm gb}$ with respect to number of particles $N$ for density $\rho=1.6$. (a)~$\phiAve$ as a function of varying $\Lol$ for various numbers of particles $N$. (b)~Coefficient of variation of $\phi_{\rm gb}$ across $\Lol$  [Eq.~\eqref{eq:cv}] for increasing values of $N$. }
\label{fig:N_scaling}
\end{figure*}

Here, we explore the scaling impact of the box geometry on the order parameter with respect to increasing number of particles for the system in the main text with $\rho=1.6$. We measure numerically the global order parameter $\phi_{\rm gb}$ for various particle numbers $N=\{32,128,528,2048,8192\}$ over 256 independent runs, and average over a total simulation time $t_{\rm max}=100$ with time step $dt=0.005$ for a range of $\Lol=[1.0,6.0]$. In short, we find the dependence of $\phiAve$ on $\Lol$ weakens as $N$ increases, with the cases $N\ge512$ showing only marginal dependence across all calculated $\Lol$ values [Fig.~\ref{fig:N_scaling}(a)]. We demonstrate this quantitatively by calculating the coefficient of variation across $\Lol$ at fixed $N$: 
\begin{equation}\label{eq:cv}
    {\rm CV} = \frac{\sqrt{\overline{\phiAve^2}-\overline{\phiAve}^2}}{\overline{\phiAve}} ,
\end{equation}
where the overbar here denotes an average over all values of $\phiAve$ for all $\Lol$ data points. Importantly, the coefficient of variation decreases with increasing $N$ [Fig.~\ref{fig:N_scaling}(b)] and is consistent with a bulk size limit where boundary dependence becomes negligible. Together, these results show that the box geometry $\Lol$ mostly affects the dynamics for small $N$ and thus should not play a major role in the limit of large $N$.


\section{Biased dynamics}
\subsection{Control interactions for bias with global order}

Adding control interactions in the dynamics is a generic strategy to improve the sampling of biased ensembles (BEs). We discuss in this section how to adapt the trajectory selection accordingly. For the unbiased dynamics
\begin{equation}
	\dot{\rvec}_i = -\mu_r \partial_{\rvec_i} V +\sqrt{2D_r} \boldsymbol\xi_i ,
	\quad
	\dot\theta_i = \omega - \mu_\theta \partial_{\theta_i} V + \sqrt{2D_\theta}\eta_i ,
	\label{eq:dynamics}
\end{equation}
the probability to observe a trajectory for $\{{\bf r}_i,\theta_i\}$ within a time interval $[0,t_o]$ is given in terms of the path weight ${\cal P}\big[\{{\bf r}_i,\theta_i\}_0^{t_o}\big] \sim e^{-{\cal A}_r-{\cal A}_\theta}$, where the actions $({\cal A}_r,{\cal A}_\theta)$ can be written in Stratonovich convention as
\begin{equation}\label{eq:action}
	\begin{aligned}
		{\cal A}_r &= \frac{1}{4 D_r} \int_0^{t_o} dt \sum_{i=1}^N ( \dot{\rvec}_i + \mu_r \partial_{\rvec_i} V )^2 - \frac{\mu_r}{2} \int_0^{t_o} dt \sum_{i=1}^N \partial_{\rvec_i}^2 V ,
		\\
		{\cal A}_\theta &= \frac{1}{4 D_\theta} \int_0^{t_o} dt \sum_{i=1}^N ( \dot\theta_i - \omega + \mu_\theta \partial_{\theta_i} V )^2 - \frac{\mu_\theta}{2} \int_0^{t_o} dt \sum_{i=1}^N \partial_{\theta_i}^2 V .
	\end{aligned}
\end{equation}
For the dynamics with the fully-connected synchronizing interaction
\begin{equation}
	\dot\theta_i = \omega - \mu_\theta \partial_{\theta_i} V + \varepsilon \sum_{j=1}^N \sin(\theta_j-\theta_i) + \sqrt{2D_\theta}\eta_i ,
	\label{eq:mod_dynamics}
\end{equation}
the probability to observe a trajectory for $\{{\bf r}_i,\theta_i\}$ within a time interval $[0,t_o]$ is given in terms of the path weight ${\cal P}_{\rm gb}\big[\{{\bf r}_i,\theta_i\}_0^{t_o}\big] \sim e^{-{\cal A}_r-{\cal A}_{\theta,\rm gb}}$, where
\begin{equation}\label{eq:action_gb}
	{\cal A}_{\theta,\rm gb} = \frac{1}{4 D_\theta} \int_0^{t_o} dt \sum_{i=1}^N \bigg[ \dot\theta_i - \omega + \mu_\theta \partial_{\theta_i} V - \varepsilon \sum_{j=1}^N \sin(\theta_j-\theta_i) \bigg]^2 - \frac{1}{2} \int_0^{t_o} dt \sum_{i=1}^N \partial_{\theta_i} \bigg[ \mu_\theta \partial_{\theta_i} V - \varepsilon \sum_{j=1}^N \sin(\theta_j-\theta_i) \bigg] .
\end{equation}
To ensure that we sample the same BE with and without control interaction, the biasing observable must be modified with respect to the control interaction. For example, for the ensemble biased with global order [Eq.~(7) in main text], we define the biasing observable $\hat\phi_{\rm gb}$ by ensuring that the biased path weight is the same with and without control:
\begin{equation}\label{eq:mod}
	{\cal P}\big[\{{\bf r}_i,\theta_i\}_0^{t_o}\big] \, e^{ - s N t_o \bar \phi_{\rm gb} } = {\cal P}_{\rm gb}\big[\{{\bf r}_i,\theta_i\}_0^{t_o}\big] \, e^{ - s N t_o \hat \phi_{\rm gb} } .
\end{equation}
Combining Eqs.~\eqref{eq:action},~\eqref{eq:action_gb} and~\eqref{eq:mod} provides an explicit expression for $\hat\phi_{\rm gb}$. Note that this procedure holds for \emph{any} value of $\varepsilon$, including $\varepsilon=0$. This grants us an additional degree of freedom with which to improve BE convergence. Here, we adopt a strategy that relates the level of chosen bias $s$ to the strength of the control force $\varepsilon$ via a relation in the observable $\bphi$. In particular for a given $s$ we choose $\varepsilon$ such that
\begin{align}
    \langle \bphi(s)  \rangle_{\rm{gb}} = \langle \bphi(\varepsilon) \rangle_{\rm{con}}~,
\end{align}
where $\langle \cdot \rangle_{\rm{gb}}$ denote the biased, controlled ensemble average and $\langle\cdot\rangle_{\rm{con}}$ the \emph{unbiased}, controlled ensemble respectively. In practice, we evaluate $\langle \bphi(\varepsilon) \rangle_{\rm{con}}$ numerically over a range of values and use the resulting function as a calibration curve with which to invert for the optimal $\varepsilon$ given a value of $s$. In the population dynamics algorithm, the update of $\varepsilon$ is implemented throughout the trajectory, until $\varepsilon$ converges to a stationary value. A similar strategy has been employed successfully in other BE studies~\cite{ABPLD1,ABPLD2}.

\begin{figure*}
  \centering
  \includegraphics[scale=.35]{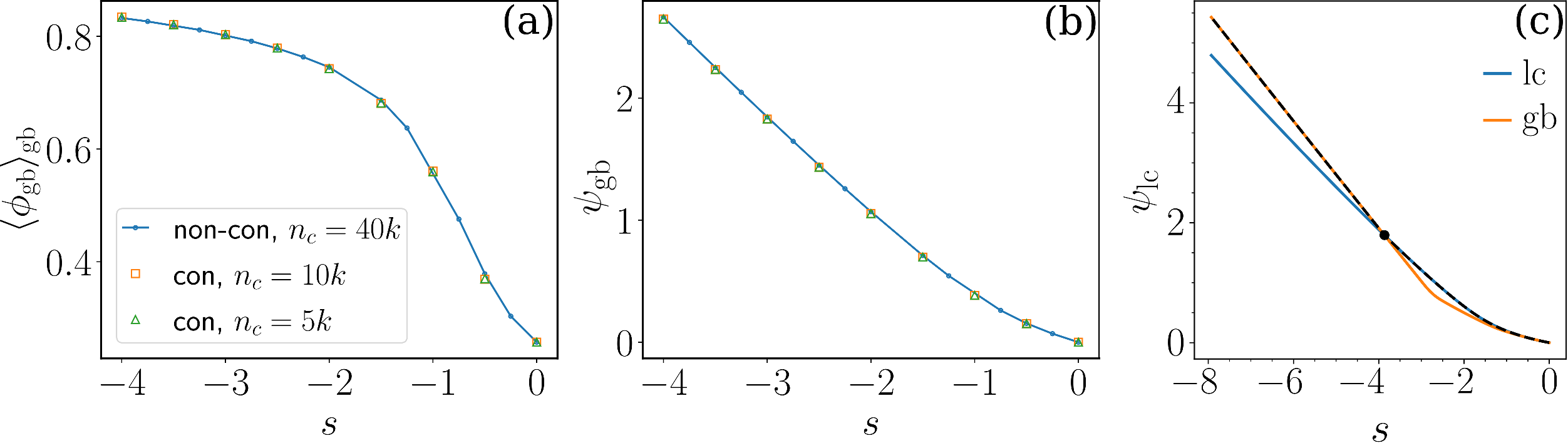}
  \caption{Ensemble biased by global order [Eq.~(7) in main text]: (a)~global order $\phiGbGb$ and (b)~scaled cumulative generating function $\psi_{\rm lc}$ as functions of the bias strength $s$.
  Blue dotted line is without control interaction and with $n_c=40000$ clones.
  Orange squares and green triangles denote data control interaction $f_c$ and with $n_c=5000$ and $n_c=10000$, respectively.
  Parameters: $N=11$, $\rho=1.2$, $t_o=16000$. 
  Ensemble biased by local order [Eq.~(10) in main text]: (c)~scaled cumulative generating function $\psi_{\rm lc}$ as a function of the bias strength $s$, sampled with either global or local control interactions in orange and blue lines, respectively. Best estimate of $\psi_{\rm{lc}}$ is shown in the dashed line.
  Parameters: $N=32$, $\rho=1.6$ and $L/\ell=1.9$
  }
\label{fig:scgf_control}
\end{figure*}

As a demonstration, we test our expression for $\hat\phi_{\rm gb}$ by comparing two numerical evaluations, with and without control, of the scaled cumulant generating function (SCGF):
\begin{equation}
	\psi_{\rm gb} = \frac{1}{Nt_o} \ln \big\langle e^{-sNt_o \bphi } \big\rangle .
\end{equation}
To this end, we consider a smaller, less dense system of $N=11$ particles at density $\rho=1.2$ than the one studied in the main text. This choice allows us to ensure convergence even without control forces, thus serving as a comparison with the results obtained with control forces. It also illustrates how control interactions accelerate sampling by reducing computational effort, here estimated by the number of clones $n_c$ used to arrive at converged solutions. As reported in Figs.~\ref{fig:scgf_control}(a-b), values of $\psi$ sampled either with or without show excellent agreement, but almost by an order of magnitude in reduced computational effort ($n_{\rm{c,noncon}}/n_{\rm{c,con}}=1/8$) with control. For all our numerical results, presented here and in the main text, we therefore use control interaction with $n_c=10000$.


\subsection{Control interactions for bias with local order}

When biasing with local order [Eq.~(10) in main text], we also consider control interaction promoting local synchronization:
\begin{equation}
	\dot\theta_i = \omega - \mu_\theta \partial_{\theta_i} V + \varepsilon \sum_{j=1}^{n_i} \sin(\theta_j-\theta_i) + \sqrt{2D_\theta}\eta_i ,
	\label{eq:mod_dynamics_2}
\end{equation}
where $n_i$ is the number of neighbors in contact with particle $i$, and for which the probability to observe a trajectory for $\{{\bf r}_i,\theta_i\}$ within a time interval $[0,t_o]$ is given in terms of the path weight ${\cal P}_{\rm lc}\big[\{{\bf r}_i,\theta_i\}_0^{t_o}\big] \sim e^{-{\cal A}_r-{\cal A}_{\theta,\rm lc}}$, where
\begin{equation}\label{eq:action_lc}
	{\cal A}_{\theta,\rm lc} = \frac{1}{4 D_\theta} \int_0^{t_o} dt \sum_{i=1}^N \bigg[ \dot\theta_i - \omega + \mu_\theta \partial_{\theta_i} V - \varepsilon \sum_{j=1}^{n_i} \sin(\theta_j-\theta_i) \bigg]^2 - \frac{1}{2} \int_0^{t_o} dt \sum_{i=1}^N \partial_{\theta_i} \bigg[ \mu_\theta \partial_{\theta_i} V - \varepsilon \sum_{j=1}^{n_i} \sin(\theta_j-\theta_i) \bigg] .
\end{equation}
As before, for either the fully-connected [Eq.~\eqref{eq:mod_dynamics}] or locally [Eq.~\eqref{eq:mod_dynamics_2}] synchronizing interactions, we define the biasing observable $\hat\phi_{\rm x}$, where $x\in\{\rm{lc},\rm{gb}\}$ respectively for each force, by ensuring that the biased path weight is the same with and without control:
\begin{align}
	{\cal P}\big[\{{\bf r}_i,\theta_i\}_0^{t_o}\big] \, e^{ - s N t_o \bar \phi_{\rm lc} } = {\cal P}_{\rm x}\big[\{{\bf r}_i,\theta_i\}_0^{t_o}\big]  
 e^{ - s N t_o \hat \phi_{\rm x} }~. 
 \end{align}
We compare the convergence for either one of the control interactions by considering the SCGF with respect to local order and the respective choice of control force:
\begin{equation}
	\psi_{\rm lc,x} = \frac{1}{Nt_o} \ln \big\langle e^{-sNt_o \bar\phi_{\rm lc} } \big\rangle_x .
\end{equation}
The SCGF $\psi_{\rm lc,x}$ may display non-convex behavior near a dynamical phase transition, which is challenging to detail numerically. Given that $\langle \bphil (s) \rangle = -\frac{d \psi(s)}{ds}$, we may also recast the expression of the SCGF as
\begin{align}
	\psi_{\rm lc,x} = -\int_{0}^{s} \langle \bphil (s') \rangle_x ds' ,
\end{align}
which is known to provide a smoother estimate of the SCGF from calculation of $\langle \bphil (s) \rangle_x$ in the algorithm~\cite{Lecomte_2007}.

Although in principle either type of interaction (i.e., local or global) provides similar results in the limit of large $t_o$ and large $n_c$, a particular choice may nonetheless best approximate the large deviation mechanism for the same amount of computational effort. This latter property is equivalent to choosing the sampling method that maximizes the value of $\psi$ for any given value of $s$. We show in Fig.~\ref{fig:scgf_control}(c) that the local interaction provides better estimates (namely higher values) of $\psi_{\rm lc}$ at small $s$, while the global interaction works best at high $s$. In the main text, Fig.~4 thus reflects the best estimate of the phase diagram given the optimal choice of the control force for a given $s$ and $L/\ell$.


\begin{figure*}
  \centering
  \includegraphics[scale=.2]{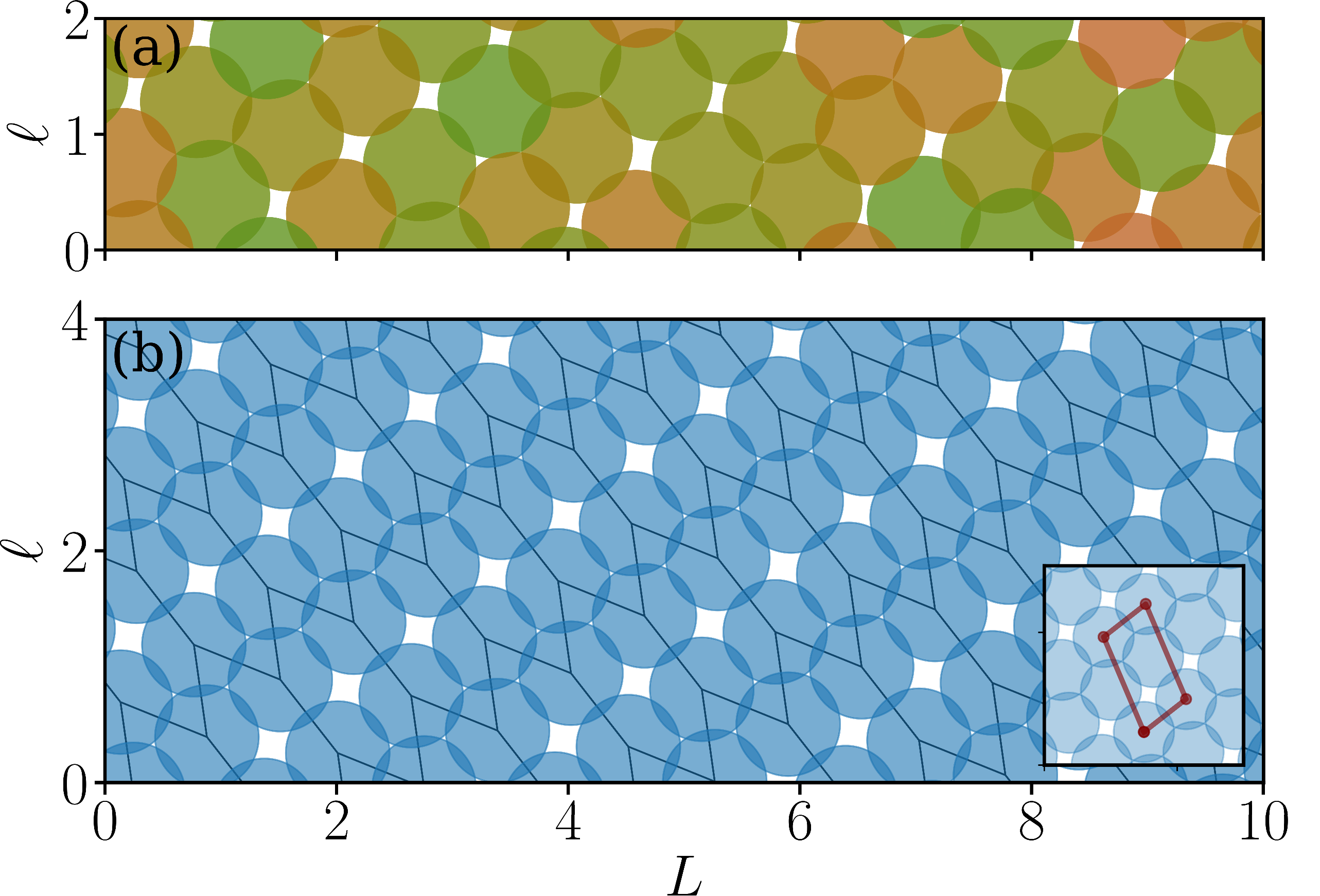}
  \caption{(a)~Biased structure for $\Lol=5$ as shown in Fig.~1(d) of the main text.
  (b)~Inferred crystal structure shown for double the range in $\ell$, and derived from the primitive lattice shown in inset. Edges in solid black lines illustrate the tiling resembling an elongated honeycomb motif.}
\label{fig:Ll5_structure}
\end{figure*}

\subsection{Packing configuration of arrest}

We characterize the ordered structure in biased dynamics at $\Lol=5$ [Fig.~1(d) of main text, here reprinted in Fig.~\ref{fig:Ll5_structure}(a)] by inferring the primitive cell that spans its underlying lattice. We find that this cell corresponds to the primitive vectors ${\bf p}_1 = (1,0)$, ${\bf p}_2 = (1/2,1+\sqrt{3}/2)$, and ${\bf b}_1=(1,1)$, where the components are dimensionalised in units of $\sigma=0.804 \sigma_0$. Generating the corresponding structure for larger system sizes [Fig.~\ref{fig:Ll5_structure}(b)], we find that the tessellation resembles a tiling by an \emph{elongated honeycomb} motif (EH). This structure may also alternatively be considered to consist of two interlocking rows of a square lattice shifted by a distance $l=\sigma/2$. Such a square motif therefore explains the higher value of $\Psi_4$ for $\Lol=5$ compared with other values of $\Lol$ [Fig.~5(a) in the main text].

Comparing the number of neighbors at the first and second coordination shells, EH displays $5$ nearest neighbors, as per the contact map results of Fig.~5(h).
It also reveals only $2$ neighbors for the next two coordination shells at $r/\sigma=\{\sqrt{2},\sqrt{3}\}$ and stands in contrast to those of the hexagonal structure---and its positions---with $6$ neighbors at $r/\sigma=\{\sqrt{3},2\}$.
In all, the excellent correspondence between the observed and generated structures provides additional support to a picture of emergent crystalline order in biased systems that yield arrest.


\subsection{Phase diagram with $N=26$ and $\rho=1.6$}

\begin{figure*}[b]
  \centering
  \includegraphics[scale=.2]{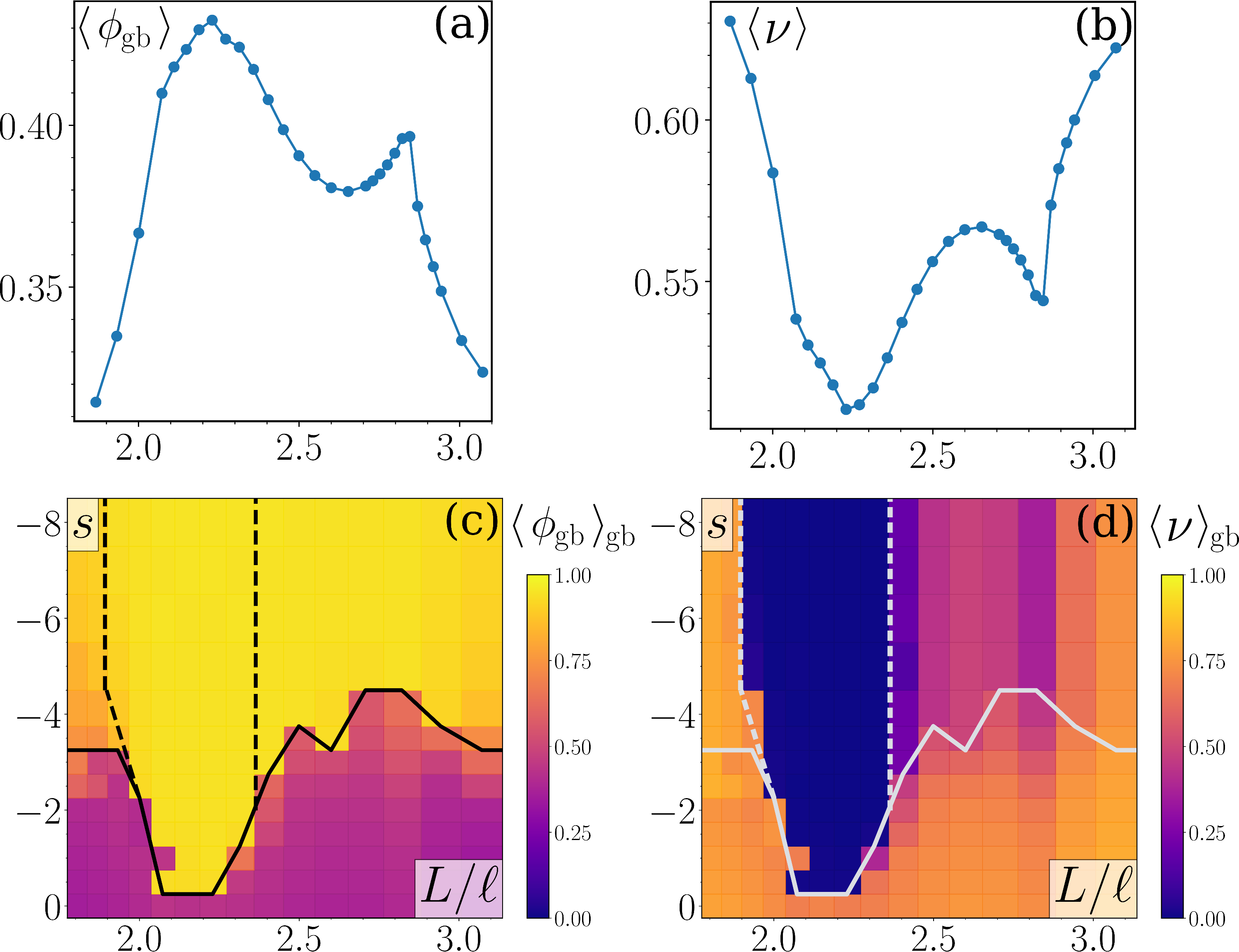}
  \caption{(a-b)~Global order $\langle \phi_{\rm gb}\rangle$ and phase current $\langle\nu \rangle$ as functions of the aspect ratio $L/\ell$ in the unbiased dynamics.
  (c-d) Phase diagram in ensembles biased by global order [Eq.~(7) in main text], in terms of global order $\phiGbGb$ and current $\nuGb$ as functions of the bias strength $s$ and the aspect ratio $\Lol$.
  Boundary lines are for $\phiGbGb = 0.65$ (solid) and $\nuGb = 0.1$ (dashed).
  Parameters: $N=26$, $\rho=1.6$.}
\label{fig:phase_N26}
\end{figure*}

To highlight the impact of box geometry on emergent dynamics, we obtain the phase diagram of a system at the same density $\rho=1.6$ as in the main text, but at a different particle number $N=26$ ($N=32$ in the main text) and box geometries $\Lol$. First, we measure in the unbiased dynamics $\langle \phi_{\rm gb}\rangle$ and $\langle\nu \rangle$ as functions of $L/\ell$ [Figs.~\ref{fig:phase_N26}(a-b)]. Increasing order systematically comes with decreasing current, as in the main text. Yet, this effect here is much sharper with distinct local maxima/minima, highlighting the highly non-monotonic effect of $\Lol$. As in the main text, regimes of relatively high order and low current at $s=0$ result in arrested states upon bias [Figs.~\ref{fig:phase_N26}(c-d)]. Moreover, we again observe for large bias a transition between cycles and arrest, when changing $\Lol$ at fixed $s$, which mirrors the slowdown of the unbiased dynamics.


\subsection{Impact of higher density on biased ensembles} 

\begin{figure*}
  \centering
  \includegraphics[scale=.32]{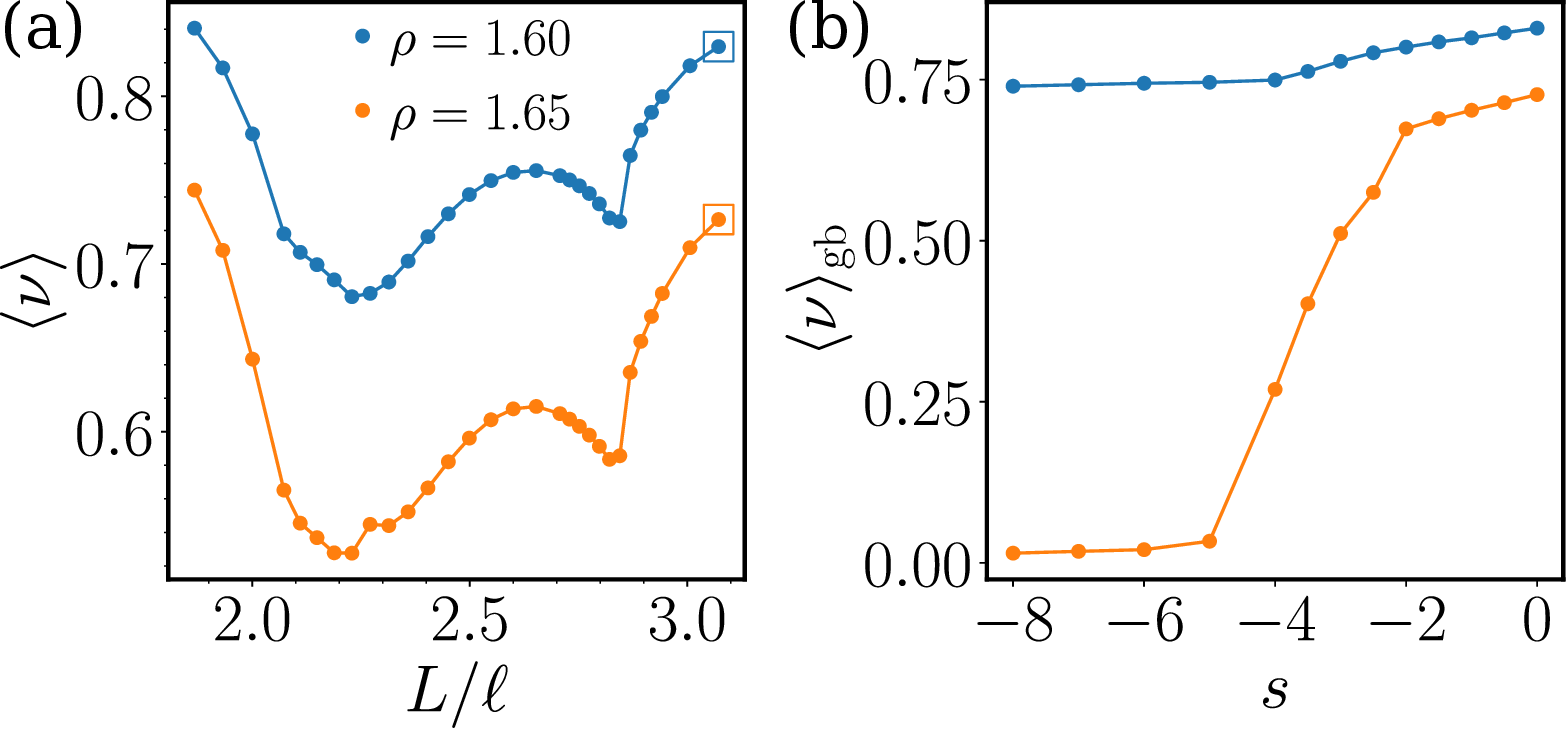}
  \caption{(a)~Unbiased average current $\nuAve$ as a function of box geometry $\Lol$ for $\rho=\{1.6,1.65\}$ respectively. (b)~Biased current average $\nuGb$ for system $\rho$ as in (a) and $\Lol=3.08$ (square markers in (a)). Parameter: $N=26$.}
\label{fig:density_analysis}
\end{figure*}

We briefly illustrate the impact of higher density on the resulting biased dynamics. In general, for the unbiased system, we expect that a higher density will generically result in higher repulsion and a lower pulsation current. Indeed, while the dependence of the current $\nuAve$ on $\Lol$ remains qualitatively similar, increasing the density systematically leads to lower $\nuAve$ [Fig.~\ref{fig:density_analysis}(a)]. Then, from the phase diagram in the biased state [Fig.~\ref{fig:phase_N26}], we expect that the biasing the higher density dynamics will likely to lead to arrest or intermittent phases. Indeed, biasing the system at $\Lol=3.08$ and $\rho=1.60$ leads to cycles, whereas biasing at $\Lol=3.08$ and $\rho=1.65$  leads to arrest as illustrated in Fig.~\ref{fig:density_analysis}(b).

Given that this choice of $\Lol$ represents a local maximum of the unbiased current $\nuAve$, we expect that any other $\Lol$ system with lower $\nuAve$ is likely to arrest under bias. Furthermore, the master curve shown in the main text [Fig.~2(c)] predicts lower currents for higher densities, suggesting that the trend in Fig.~\ref{fig:density_analysis} should hold generically for other density values.

%